\documentclass[useAMS,usenatbib,aas_macros]{mn2e}

\usepackage{epsfig}
\usepackage{lscape,graphicx}
\usepackage{amssymb}
\usepackage{natbib}
\usepackage{times}
\usepackage{aas_macros}

\newcommand{\abs}[1]{\left\vert#1\right\vert}

\newcommand{\msun}{M_\odot}
\newcommand{\ifm}[1]{\relax\ifmmode#1\else$\mathsurround=0pt #1$\fi}
\newcommand{\hmpc}{\,\ifm{h^{-1}}{\rm Mpc}}
\newcommand{\hmsun}{\,\ifm{h^{-1}}{M_{\odot}}}
\def\kms{\rm \; km \; sec^{-1}}

\newcommand{\equ}[1]{eq.~(\ref{eq:#1})}

\newcommand{\se}[1]{\S\ref{sec:#1}}

\newcommand{\dd}{{\rm d}}
\newcommand{\dS}{\Delta S}
\newcommand{\dW}{\Delta \omega}

\newcommand{\etal}{{et al.~}}
\newcommand{\K}{\>{\rm K}}

\def\mbig{M_{\rm main}}
\def\mall{M_{\rm all}}
\def\bmbig{\bar{M}_{\rm main}}
\def\bmall{\bar{M}_{\rm all}}
\def\zbig{z_{\rm main}}
\def\bzbig{\bar{z}_{\rm main}}
\def\zall{z_{\rm all}}
\def\bzall{\bar{z}_{\rm all}}
\def\wbig{\omega_{\rm main}}
\def\bwbig{\bar{\omega}_{\rm main}}

\def\bwall{\bar{\omega}_{\rm all}}
\def\mmin{M_{\rm min}}
\def\smin{S_{\rm min}}
\def\bmbiga{\bar{M}_{\rm main,1}}

\def\bwbiga{\bar{\omega}_{\rm main,1}}
\def\dmain{D_{\rm main}}
\def\dall{D_{\rm all}}
\def\ltsima{$\; \buildrel < \over \sim \;$}
\def\lsim{\lower.7ex\hbox{\ltsima}}
\def\lta{\la}

\begin{document}


\title[Natural Downsizing in Hierarchical Galaxy Formation]
      {Natural Downsizing in Hierarchical Galaxy Formation}

\author[E.~Neistein, F.~C.~van den Bosch, \& A.~Dekel]
       {Eyal Neistein$^1$,
        Frank C. van den Bosch$^2$,
        and Avishai Dekel$^1$\\
     $^1$ Racah Institute of Physics, The Hebrew University,
          Jerusalem, Israel\\
     $^2$ Max-Planck-Institute for Astronomy, K\"onigstuhl 17, D-69117
          Heidelberg, Germany\\
     e-mails: eyal$\underline{\;\;}$n@phys.huji.ac.il;
              vdbosch@mpia-hd.mpg.de;
              dekel@phys.huji.ac.il }


\date{}
\pagerange{\pageref{firstpage}--\pageref{lastpage}} \pubyear{2006}
\maketitle

\label{firstpage}


\begin{abstract}
  Stellar-population  analyses of today's galaxies show ``downsizing",
  where the stars in more massive galaxies tend to have formed earlier
  and  over  a  shorter time span. We show that this phenomenon is not
  necessarily ``anti-hierarchical" but rather has its natural roots in
  the  bottom-up  clustering  process of dark-matter haloes. While the
  main  progenitor does indeed show an opposite effect, the integrated
  mass  in  all  the  progenitors down to a given minimum mass shows a
  robust  downsizing  that  is  qualitatively similar to what has been
  observed.  These  results are derived analytically from the standard
  extended  Press  Schechter (EPS) theory, and are confirmed by merger
  trees   based  on  EPS  or  drawn  from  $N$-body  simulations.  The
  downsizing  is valid for any minimum mass, as long as it is the same
  for  all  haloes  at  any  given  time, but the effect is weaker for
  smaller  minimum  mass.  If efficient star formation is triggered by
  atomic  cooling,  then a minimum halo mass arises naturally from the
  minimum  virial  temperature  for  cooling, $T \simeq 10^4$K, though
  for  such  a  small minimum mass the effect is weaker than observed.
  Baryonic  feedback  effects,  which  are  expected  to  stretch  the
  duration  of  star  formation  in small galaxies and shut it down in
  massive  haloes  at  late  epochs,  are  likely to play a subsequent
  role   in   shaping   up   the  final  downsizing  behaviour.  Other
  appearances  of  downsizing,  such  as  the decline with time of the
  typical  mass of star-forming galaxies, may not be attributed to the
  gravitational  clustering  process  but  rather  arise  from the gas
  processes.
\end{abstract}




\section{Introduction}
\label{sec:intro}

A key issue  in the study of galaxy  formation is the anti-correlation
between the  stellar mass of a  galaxy and the formation  epoch of the
stars in it,  which is referred to in general  terms as ``downsizing".
In its most  pronounced form, this is simply  the fact that elliptical
galaxies  consist of  old  stellar  populations and  tend  to be  more
massive while disc galaxies have  younger stars and are typically less
massive.  However, a similar  correlation between stellar mass and age
is detected within each of  the two major classes of galaxies, whether
they are  classified morphologically as ellipticals  versus spirals or
by colour  as ``red sequence" versus ``blue  sequence" galaxies. These
trends are quite robust, e.g.,  they are insensitive to how luminosity
is translated to stellar mass and colour to stellar age.

A downsizing effect can actually appear in different forms which refer
to  different phenomena,  involving  different types  of galaxies  and
different  epochs in  their histories.   One form,  which is  the main
focus  of the  current  paper, is  the  fact that  the star  formation
histories inferred  from present-day galaxies  using synthetic stellar
evolution models  correlate with galactic  stellar mass. The  stars in
more massive  galaxies tend  to form  at an earlier  epoch and  over a
shorter  time   span.  This  phenomenon   is  termed  ``archaeological
downsizing"  \citep[ADS, following][]{Thomas05}.  Using  observed line
indices  and abundance  ratios, ADS  has been  detected  in elliptical
galaxies \citep{Thomas05,Nelan05},  and in a large  sample of galaxies
from the Sloan Digital Sky Survey \citep{Heavens04,Jimenez05}.

The other face of downsizing is the fact that the sites of active star
formation shift from  high-mass galaxies at early times  to lower mass
systems  at lower redshift.  We term  this phenomenon  ``downsizing in
time" (DST).  It has first been detected by \citet{Cowie96}, who found
that the maximum rest-frame $K$-band luminosity of galaxies undergoing
rapid  star formation  has been  declining smoothly  with time  in the
redshift range $z=0.2-1.7$.  This DST phenomenon has been confirmed by
numerous                       subsequent                      studies
\citep{Guzman97,Brinchmann00,Kodama04,Juneau05, Bell05, Bundy05}.

It is important  to realize that these two forms  of downsizing can be
very  different, and  possibly  even orthogonal  to  each other.   The
archaeological  analysis of  local galaxies  highlights  the formation
epoch of  the majority of their stars,  which at least in  the case of
ellipticals  occurs at  high  redshifts, $z  \sim  2-5$. In  contrast,
downsizing in  time refers to the specific  star-formation rate (SSFR)
at  relatively low  redshifts, $z  \lta 1$,  and therefore  focuses on
later  phases  of star  formation,  which  may  involve only  a  small
fraction of the stars in  the galaxy.  Unless the stellar-mass ranking
of present  day galaxies is the  same as that of  their progenitors at
higher  redshifts, these two  forms of  downsizing do  not necessarily
reflect the same phenomenon.  Since hierarchical clustering in general
does  not preserve  this mass  ranking,  the two  forms of  downsizing
should  be treated  as two  different phenomena.   Indeed, as  we will
demonstrate  below,  the  understanding  of  one  does  not  imply  an
understanding of the other.

In the standard $\Lambda$CDM cosmological scenario, dark-matter haloes
are  built hierarchically bottom up. This is obvious for the evolution
of   individual  haloes,  as  they  are  constructed  by  the  gradual
gravitational   assembly   of  smaller  progenitor  haloes  that  have
collapsed  and  virialized  earlier  on.  The bottom-up clustering can
also  be  inferred  statistically  from  the power spectrum of initial
density  fluctuations,  which  indicates that the mass distribution of
collapsing  systems  is  shifting  in time from small to large masses.
These  hierarchical  aspects of the clustering process have led to the
misleading  notion  that  one expects big haloes to ``form" later than
small haloes, without distinguishing between the dynamical collapse or
assembly of these haloes and the formation epoch of the stars in them.
The  observed  downsizing  is  therefore frequently referred to in the
literature  as  ``anti-hierarchical",  and  thus  as  posing  a severe
challenge  to  the  standard  model  for structure formation. However,
when  comparing  the  histories of different haloes, the evolution may
be   interpreted   as   bottom-up   or   top-down   depending  on  how
``formation"  is  defined.

The  evolution  of dark  matter  (DM)  haloes  has traditionally  been
studied  through   the  histories   of  the  {\it   main  progenitors}
\citep{Lacey93,   Eisenstein96,   Nusser99,   Firmani00,   vdBosch02b,
  Wechsler02,   Li05}.   The   main-progenitor  assembly   history  is
constructed by following  back in time the most  massive progenitor in
each  merger event.  We  term $\mbig(z)$  the main progenitor  mass at
redshift $z$.  The corresponding  formation redshift $\zbig$ of a halo
of mass  $M_0$ at $z=0$ is commonly  defined as the time  at which the
main    progenitor    contained   one    half    of   today's    mass,
$\mbig(\zbig)=M_0/2$.   According  to  this definition,  more  massive
haloes   indeed  form  later.  The  formation  redshift  of  the  main
progenitor  has been computed by \citet{Lacey93} based on the Extended
Press-Schechter  (EPS)  formalism,  and  the  trend with mass has been
confirmed  for  various  cosmologies \citep[see for example][hereafter
vdB02]{vdBosch02b}.  This  has  also been tested using trees extracted
from  cosmological  $N$-body  simulations \citep{Lacey94, Wechsler02}.
We  confirm  this  behaviour  below  using  a  new analytic estimation
of  the  full time evolution of $\mbig(z)$, based on the EPS formalism
itself   without  the  need  to  construct  merger-tree  realizations.

However, the history  of the main progenitor of a  given halo does not
represent the history of the  whole population of progenitors in which
the stars  of a present-day  halo have formed.  Perhaps  more directly
relevant for the stellar population at any given epoch is the sum over
the masses  of {\it all}  the virialized progenitors in  that specific
tree  at  that  time,  which  we term $\mall(z)$. If this summation is
performed  down  to  a  zero  minimum  mass,  we  have  by  definition
$\mall(z)=M_0$.  However,  when  a non-zero minimum mass $\mmin(z)$ is
applied,  the  same  for  all  haloes,  we find a robust archeological
downsizing  behaviour.\footnote{a  similar  point  has  been  made  in
  parallel  by \citet{Mouri06} using a very different methodology.} We
demonstrate  this  effect  analytically based on the EPS formalism and
confirm  it  using  Monte-Carlo  EPS  merger  trees  as  well as trees
extracted  from $N$-body simulations. We prove that this phenomenon is
valid  for  any  realistic  power-spectrum shape and for any choice of
$\mmin(z)$,  as long as it is the same for all haloes at a given time.
We  note that a similar trend has been found by \citet{Bower91} for an
Einstein-deSitter cosmology and a power-law power spectrum.

\begin{figure}
\centerline{ \hbox{ \epsfig{file=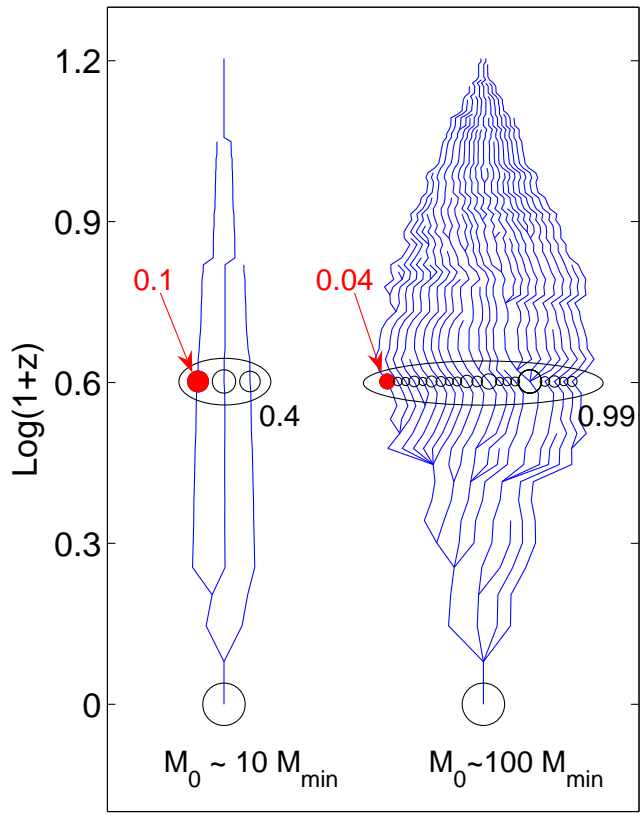,width=9cm} } } \caption{An
illustration of the upsizing of $\mbig$ versus the downsizing of
$\mall$ in dark-halo merger trees. Compared are random trees drawn
from the EPS probabilities for haloes of current masses $M_0 \sim
10\mmin$ and $\sim 100\mmin$. The mass of the main progenitor versus
the total mass in all the progenitors above $\mmin$ are shown at
$z=3$. The progenitors, of mass $M$, are marked by circles of sizes
and spacings proportional to $(M/M_0)^{1/3}$. The values of
$\mbig/M_0$ and $\mall/M_0$ are indicated. The main progenitor, along
the left branch, is more massive in the less massive $M_0$, showing
upsizing. The integrated mass in all the progenitors down to $\mmin$
is larger in the massive $M_0$, demonstrating ``downsizing".}
  \label{figs:ds_tree}
\end{figure}

\begin{figure}
\centerline{ \hbox{ \epsfig{file=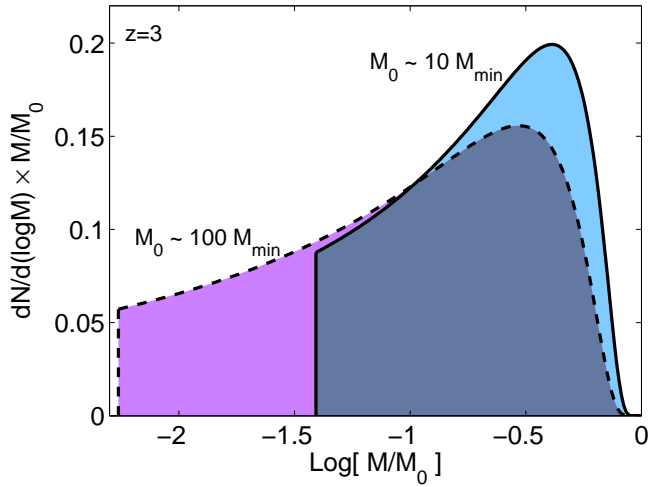,width=9cm} } }
\caption{Upsizing of $\mbig$ versus downsizing of $\mall$ in the
distribution of mass in progenitors at $z=3$. The area under each
curve, from $\log(M/M_0)$ to 0, is the total mass in progenitors above
$M$ relative to $M_0$. The excess of mass in massive progenitors for
the smaller current halo indicates upsizing of $\mbig$. The excess in
total mass down to $\mmin/M_0$ for the more massive current halo
demonstrates downsizing of $\mall$.}
  \label{figs:dndm}
\end{figure}

The difference between $\mbig$ and $\mall$ is illustrated in
Fig.~\ref{figs:ds_tree}, which compares the $z=3$ progenitors above a
given $\mmin$ in random realizations of merger trees corresponding to
current haloes of $M_0 \sim 10\mmin$ and $\sim 100\mmin$. The
downsizing behaviour for $\mall$ is apparent, while for $\mbig$ the
familiar opposite trend stands out (we term this trend as
``upsizing'').
The average distributions of relative masses in $z=3$ progenitors,
derived using EPS (see below) for the same two values of $M_0$ as in
Fig.~\ref{figs:ds_tree}, are shown in Fig.~\ref{figs:dndm}. The
upsizing of $\mbig$ is indicated by the excess of massive progenitors
for the smaller current halo. The downsizing of $\mall$ is
demonstrated by the excess of the overall integral down to $\mmin/M_0$
for the more massive current halo.

A realistic and necessary condition for star formation is that the gas
is able to cool efficiently. This  is only possible if the gas resides
in a  halo whose  virial temperature exceeds  a critical  threshold of
$T\sim 10^4$K, above which  atomic cooling becomes efficient. This
provides a  natural threshold  $\mmin(z)$ for  $\mall(z)$. If star
formation is of the maximum possible efficiency, namely if all the gas
in haloes  above $\mmin(z)$ turns  into stars on  a free-fall
time-scale, then  the ADS in the stellar  population emerges naturally
from the ADS of $\mall(z)$.

In reality,  however, the star formation  rate is likely  to be slowed
down  by a variety  of baryonic  processes, especially  by ``feedback"
effects.   As a  result, the  star-formation  history may  or may  not
maintain the ADS seeded by $\mall(z)$ of the DM haloes. This should in
principle  be  modeled  by   semi-analytic  models  (SAMs)  of  galaxy
formation,  which   attempt  to  incorporate   the  baryonic  physical
processes  in merger trees  of DM  haloes.  Unfortunately,  early SAMs
failed  to  reproduce the  ADS  of ellipticals  as  we  know it  today
\citep[e.g.,   ][]{Kauffmann96,    Baugh96,   Kauffmann98,   Thomas99,
  Thomas99a},  probably due  to  an inadequate  treatment of  feedback
effects.  SAMs also failed to  recover the similar global trend obeyed
by     blue-sequence    galaxies    in     color-magnitude    diagrams
\citep{vdBosch02a, Bell03}, thus highlighting the apparent discrepancy
between   theory  and  observation.    However,  more   recent  models
\citep[e.g., ][]{Bower05, DeLucia06,  Croton06, Cattaneo06} do succeed
in  reproducing  an ADS  behaviour,  largely  because  of an  improved
treatment  of the  feedback  effects.  The  early  SAMs only  included
supernova feedback, which is  efficient in slowing down star formation
preferentially in  smaller galaxies below  a virial velocity  of $\sim
100\kms$ \citep{Dekel86}. The problem is that this process only causes
a {\it  delay} in the star  formation: the gas is  only prevented from
forming  stars until  the  halo has  grown  sufficiently massive  that
supernova feedback  is no longer  efficient.  Because this  results in
relatively  late star formation,  even in  massive galaxies,  the SAMs
were unable to predict the  correct stellar ages. The main success of
the more modern SAMs is the  inclusion of AGN feedback and shock
heating physics, which causes a shutdown, rather  than a delay,  of
star formation at relatively late times \citep[e.g.,  ][]{Birnboim03,
Binney04, Croton06, Scannapieco05, Cattaneo06, Dekel06}. Although the
details of AGN feedback are still poorly understood, it has been
argued   that it is   the main mechanism that explains   the
``anti-hierarchical'' nature of the relation between stellar mass and
stellar age of galaxies.

However, we show below that  the simulated star formation histories of
elliptical galaxies \citep{DeLucia06} are qualitatively similar to the
histories  predicted  by  $\mall(z)$   of  dark  matter  haloes.  This
indicates that the  roots of the observed ADS can  be found already in
the  natural downsizing of  the dark  matter haloes.   Apparently, the
complex feedback processes affecting  the star formation do not change
the general trend and only  provide fine-tuning to the ADS effect.  We
conclude   that  ADS   should  not   be  regarded   as   a  surprising
``anti-hierarchical"  phenomenon  of complex  gas  physics  --- it  is
rather  the  most  natural,  expected behaviour  in  the  hierarchical
clustering scenario.

On  the other hand, we find that the downsizing in time as observed at
relatively  low  redshifts  cannot  be  easily traced back to the bare
properties  of  the dark matter merger trees. The mass distribution of
late-type  efficient  star  formers  at  late  times  must be strongly
affected by feedback or other gas processes and therefore the modeling
of  this  aspect  of  downsizing  should  involve  more realistic star
formation  rates.  We  show  that  only  when  $\mmin(z)$  is properly
increasing  with  redshift,  possibly  mimicking the required baryonic
effects,  the  star-formation  rate  associated with $\mall(z)$ can be
forced  to  a  qualitative  agreement  with the observed downsizing in
time.

The  paper is  organized as  follows.  In  the  following introductory
section,  \se{EPS}, we  spell  out  the relevant  items  from the  EPS
formalism and  describe how we generate Monte-Carlo  merger trees that
serve  us as a  reference when  needed. In  \se{mbig}, we  address the
average  $\mbig(z)$,  derive an  analytic  approximation  for it,  and
confirm that it behaves opposite  to downsizing. In \se{mall} we study
the average $\mall(z)$, compute it analytically from the  EPS
formalism,  and demonstrate that  it shows a  robust ADS behaviour. We
also study  the mutual correlation between the formation times
associated with $\mbig$  and $\mall$.  In \se{formation_rate} we
compute  the EPS  formation rate  of DM  haloes of  a given  mass, and
compare it with  star-formation histories in semi-analytic simulations
and in observations.  In \se{TDS} we address the downsizing in time of
the SSFR as observed at different redshifts out to $z\sim 1$. In
\se{discuss} we summarize our results and discuss them.

Throughout this paper we use a flat $\Lambda$CDM cosmology, with the
standard power spectrum $P(k)=kT^2(k)$. The transfer function
\citep{Bardeen86} is
\begin{eqnarray}
\lefteqn{T(k) = \frac{\ln(1+2.34q)}{2.34q}\times } \\
\nonumber & & \left[1 + 3.89q + (16.1q)^2 + (5.46q)^3 + (6.71q)^4
\right] ^{-1/4} \,.
\end{eqnarray}
Here $q = k/\Gamma$, with $k$ in $h$Mpc$^{-1}$, and $\Gamma$ is the
power spectrum shape parameter \citep{Sugiyama95}
\begin{equation}
\Gamma = \Omega_m h \exp \left[ -\Omega_b(1+\sqrt{2h}/\Omega_m)
\right] \,,
\end{equation}
where $\Omega_b=0.044$ throughout the paper. Unless specifically
stated otherwise, we use the standard cosmological parameters, with
$\Omega_{\Lambda} =0.7$, $\Omega_{\rm m}=0.3$, $\sigma_8=1.0$,  and
$h=0.7$ (whenever we modify $\Omega_m$ or $h$ we recompute $\Gamma$
according to the above definition).

\section{Extended Press-Schechter theory}
\label{sec:EPS}

\subsection{The Formalism}
\label{sec:EPSformalism}

In  the  standard model  for  structure  formation  the initial  density
contrast $\delta({\bf x}) =  \rho({\bf x})/\bar{\rho} - 1$ is considered
to be a  Gaussian random field, which is  therefore completely specified
by the power  spectrum $P(k)$.  As long as $\delta \ll  1$ the growth of
the  perturbations  is linear  and  $\delta({\bf  x},t_2) =  \delta({\bf
  x},t_1)  D(t_2)/D(t_1)$, where  $D(t)$ is  the linear  growth  factor.
Once  $\delta({\bf x})$  exceeds a  critical  threshold $\delta^{0}_{\rm
  crit}$ the perturbation starts to collapse to form a virialized object
(halo).  In the case of spherical collapse $\delta^{0}_{\rm crit} \simeq
1.68$.   In what  follows we  define $\delta_0$  as the  initial density
contrast field linearly  extrapolated to the present time.   In terms of
$\delta_0$, regions  that have collapsed  to form virialized  objects at
redshift $z$ are then associated  with those regions for which $\delta_0
> \delta_c(z) \equiv \delta^{0}_{\rm crit}/D(z)$.

In   order  to   assign   masses  to   these   collapsed  regions,   the
Press-Schechter (PS) formalism considers the density contrast $\delta_0$
smoothed with a spatial window function (filter) $W(r;R_f)$.  Here $R_f$
is a characteristic size of the  filter, which is used to compute a halo
mass $M = \gamma_f \bar{\rho}  R_f^3/3$, with $\bar{\rho}$ the mean mass
density of the Universe and $\gamma_f$ a geometrical factor that depends
on  the  particular  choice of  filter.   The  {\it  ansatz} of  the  PS
formalism is that the fraction of mass that at redshift $z$ is contained
in  haloes with  masses  greater than  $M$  is equal  to  two times  the
probability that  the density contrast smoothed  with $W(r;R_f)$ exceeds
$\delta_c(z)$.  This results in the  well known PS mass function for the
comoving number density of haloes:
\begin{eqnarray}
\label{eq:PS}
\lefteqn{{{\dd}n \over {\dd} \, {\rm ln} \, M}(M,z) \, {\dd}M =}
\nonumber \\ & & \sqrt{2 \over \pi} \, \bar{\rho} \, {\delta_c(z)
\over \sigma^2(M)} \, \left| {{\dd} \sigma \over {\dd} M}\right| \,
{\rm exp}\left[-{\delta_c^2(z) \over 2 \sigma^2(M)}\right] \, {\dd}M
\end{eqnarray}
\citep{Press74}.  Here $\sigma^2(M)$ is the mass variance of the
smoothed density field given by
\begin{equation}
\label{eq:variance} \sigma^2(M) = {1 \over 2 \pi^2}
\int_{0}^{\infty} P(k) \; \widehat{W}^2(k;R_f) \; k^2 \; {\dd}k \,,
\end{equation}
with $\widehat{W}(k;R_f)$  the Fourier transform  of $W(r;R_f)$.

The   {\it   extended}   Press-Schechter   (EPS)  model  developed  by
\citet{Bond91},  is  based  on  the  excursion set formalism. For each
point  one constructs `trajectories' $\delta(M)$ of the linear density
contrast  at  that  position as function of the smoothing mass $M$. In
what   follows   we   adopt   the  notation  of  \citet[][ hereafter
LC93]{Lacey93}  and  use the variables $S = \sigma^2(M)$ and $\omega =
\delta_c(z)$  to  label  mass and redshift, respectively. In the limit
$R_f  \rightarrow  \infty$ one has that $S = \delta(S) = 0$, which can
be  considered  the starting point of the trajectories. Increasing $S$
corresponds  to  decreasing  the  filter  mass  $M$,  and  $\delta(S)$
starts  to  wander  away  from  zero,  executing a random walk (if the
filter  is  a  sharp  $k$-space  filter).  The  fraction  of matter in
collapsed  objects  in the mass interval $M$, $M+{\rm d}M$ at redshift
$z$  is  now  associated  with  the fraction of trajectories that have
their   {\it   first   upcrossing}   through  the  barrier  $\omega  =
\delta_c(z)$  in  the  interval  $S$,  $S+{\rm d}S$, which is given by
\begin{equation}
\label{eq:probS}
f(S ,\omega) \; {\dd}S = {1  \over \sqrt{2 \pi}} \;
{\omega  \over S^{3/2}} \; {\rm exp}\left[-{\omega^2 \over 2
S}\right] \; {\dd}S
\end{equation}
\citep[][   LC93]{Bond91,  Bower91}.    After  conversion   to  number
counting,  this probability function  yields the  PS mass  function of
equation~(\ref{eq:PS}).  Note that this  approach does not suffer from
the arbitrary factor two in the original Press \& Schechter approach.

Since for random walks the upcrossing probabilities are independent of
the  path  taken (i.e.,  the  upcrossing  is  a Markov  process),  the
probability for a change $\dS$ in a time step $\dW$ is simply given by
equation~(\ref{eq:probS})  with $S$ and  $\omega$ replaced  with $\dS$
and $\dW$, respectively.  This allows  one to immediate write down the
{\it conditional} probability that a  particle in a halo of mass $M_2$
at $z_2$ was  embedded in a halo  of mass $M_1$ at $z_1$  (with $z_1 >
z_2$) as
\begin{eqnarray}
\label{eq:probSS}
\lefteqn{P(S_1,\omega_1 \vert  S_2,\omega_2) \; {\dd}S_1 =
f(S_1-S_2,\omega_1-\omega_2)\dd S_1 = } \nonumber \\
& & {1  \over \sqrt{2 \pi}} \; {(\omega_1    -   \omega_2)    \over
(S_1    -    S_2)^{3/2}} \; {\rm exp}\left[-{(\omega_1 - \omega_2)^2
\over 2 (S_1 - S_2)}\right] \; {\dd}S_1 \,.
\end{eqnarray}
Converting from  mass weighting to  number weighting, one  obtains the
average number  of progenitors  at $z_1$ in  the mass  interval $M_1$,
$M_1 + {\rm d}M_1$ which by  redshift $z_2$ have merged to form a halo
of mass $M_2$:
\begin{eqnarray}
\label{eq:condprobM}
\lefteqn{{{\dd}N \over {\dd}M_1}(M_1,z_1 \vert M_2,z_2) \; {\dd}M_1 =}
\nonumber \\
& & {M_2 \over M_1} \; P(S_1,\omega_1 \vert S_2,\omega_2) \;
\left\vert {{\dd}S \over {\dd M}} \right\vert \; {\dd}M_1 \,.
\end{eqnarray}
%

\subsection{Constructing Merger-Trees}
\label{sec:merger_trees}

The  conditional  mass  function  can  be  combined  with  Monte-Carlo
techniques to construct merger histories (also called merger trees) of
dark matter  haloes.  If  one wants to  construct a set  of progenitor
masses  for  a  given  parent   halo  mass,  one  needs  to  obey  two
requirements.   First,  the {\it  number}  distribution of  progenitor
masses   of   many    independent   realizations   needs   to   follow
(\ref{eq:condprobM}). Second,  mass needs to be conserved,  so that in
each individual realization the sum  of the progenitor masses is equal
to the  mass of the parent  halo.  In principle,  this requirement for
mass conservation  implies that  the probability for  the mass  of the
$n^{\rm th}$ progenitor  needs to be conditional on  the masses of the
$n-1$   progenitor  haloes   already   drawn.   Unfortunately,   these
conditional probability  functions are unknown, and one  has to resort
to an approximate technique for the construction of merger trees.

The  most  widely adopted algorithm is the $N$-branch tree method with
accretion  developed  by \citet[][ hereafter SK99]{Somerville99}. This
method  is  more  reliable  than for example the binary-tree method of
LC93.   In   particular,  it  ensures  exact  mass  conservation,  and
yields  conditional  mass  functions  that  are in good agreement with
direct   predictions   from   EPS   theory   (i.e.,   the   method  is
self-consistent).

The SK99  method works as  follows.  First a  value for $\Delta  S$ is
drawn from the mass-weighted probability function
\begin{equation}
\label{probdS}
f(\dS ,\dW)  \; {\dd}\dS = {1 \over \sqrt{2  \pi}} \;
{\dW \over \dS^{3/2}} \; {\rm exp}\left[-{(\dW^2) \over 2
\dS}\right] \; {\dd}\dS
\end{equation}
[cf.  equation~(\ref{eq:probSS})].  Here  $\dW$ is  a measure  for the
time  step used  in the  merger  tree, and  is a  free parameter  (see
below).  The  progenitor mass,  $M_p$, corresponding to  $\dS$ follows
from $\sigma^2(M_p) = \sigma^2(M) + \dS$.  With each new progenitor it
is checked  whether the  sum of the  progenitor masses drawn  thus far
exceeds  the  mass  of the  parent,  $M$.  If  this  is the  case  the
progenitor  is  rejected and  a  new  progenitor  mass is  drawn.  Any
progenitor with  $M_p <  M_{\rm min}$ is  added to the  mass component
$M_{\rm acc}$ that  is considered to be accreted onto  the parent in a
smooth  fashion  (i.e., the  formation  history  of  these small  mass
progenitors is not followed further  back in time). Here $M_{\rm min}$
is a  free parameter that has  to be chosen  sufficiently small.  This
procedure is repeated  until the total mass left, $M_{\rm  left} = M -
M_{\rm acc} -  \sum M_p$, is less than  $M_{\rm min}$.  This remaining
mass is  assigned to $M_{\rm acc}$ and  one moves on to  the next time
step.

As  all  other  methods  for  constructing  merger  trees \citep[e.g.,
LC93;  ][]{Kauffmann93},  the SK99 algorithm is only an approximation.
In  particular,  it  is  based  on  the {\it mass}-weighted progenitor
probability  function  (\ref{probdS}), rather than on the {\it number}
distribution  (\ref{eq:condprobM}),  and mass conservation is enforced
`by  hand',  by  rejecting  progenitor  masses  that overflow the mass
budget.  Consequently,  the  number  distribution  of  the first-drawn
progenitor   masses   is  different  from  that  of  the  second-drawn
progenitor  masses,  etc.  Somewhat  fortunately,  the  sum  of  these
distributions     closely     matches    the    number    distribution
(\ref{eq:condprobM})  of  all  progenitors,  but  only if sufficiently
small  time  steps  $\dW$ are used (see SK99 and vdB02). In principle,
since  the  upcrossing  of random walks through a boundary is a Markov
process,  the statistics of progenitor masses should be independent of
the  time  steps  taken,  indicating  that the method is not perfectly
justified.  Consequently,  not all statistics of the merger trees thus
constructed are necessarily accurate, something that has to be kept in
mind.

In this paper we adopt a time step of
\begin{equation}
\dW = \sqrt{ \left|\frac{\dd S}{\dd M}\right| 10^{-3}M } \left[ b +
a\log_{10} \left( \frac{M}{M_{\rm min}} \right) \right] ^{-1} \,,
\end{equation}
where $a=0.3$, $b=0.8$. As shown in SK99, this time step yields number
distributions  of progenitor masses  that are  in good  agreement with
(\ref{eq:condprobM}). The average number  of progenitors per time step
is $\sim 1.5$ for $10^9 h^{-1}\msun \leq M \leq 10^{14} h^{-1}\msun$
and $M_{\rm min}=10^8 h^{-1} \msun$.

\section{Growth of the Main Progenitor}
\label{sec:mbig}

The  full  merger  history  of  any  individual  dark matter halo is a
complex  structure  containing  a lot of information. It has therefore
been  customary  to  define a {\it main progenitor history}, sometimes
termed   {\it   mass   accretion  history}  \citep[][vdB02]{Firmani00,
Wechsler02}   or  {\it  mass  assembly  history}  \citep{Li05},  which
restricts attention to the main ``trunk" of the merger tree. This main
trunk  is  defined  by  following  the branching of a merger tree back
in  time,  and  selecting  at  each  branching  point the most massive
progenitor.  We  denote by $\mbig(z)$ the mass of this main progenitor
as  a  function  of  redshift  $z$.  Note  that  with this definition,
the  main progenitor is not necessarily the most massive progenitor of
its  generation  at  a  given time, eventhough it never accretes other
haloes that are more massive than itself.

\subsection{Analytical Derivation}
\label{sec:mbig_sub}

Using  EPS  merger  trees and cosmological $N$-body simulations, vdB02
and  \citet{Wechsler02}  have obtained simple fitting formulae for the
main   progenitor   history.  We  show  here  that  one  can  actually
derive   a   useful   analytical   approximation   for   the   average
$\bmbig(z)$,   defined  at  each  redshift  as  the  average  mass  of
$\mbig(z)$  over  an  ensemble  of merger-trees. We derive it directly
from  the  EPS  formalism,  without  the need to construct Monte-Carlo
merger  trees.  As  shown  in  the  Appendix,  $\bmbig(z)$  obeys  the
differential equation
\begin{equation}
\label{eq:mbig_de} \frac{\dd\bmbig}{\dd\omega}
=-\sqrt{\frac{2}{\pi}}\frac{\bmbig}{\sqrt{S_q-S}} \, .
\end{equation}
Here $S=S(\bmbig)$, $S_q=S(\bmbig/q)$, and the value of $q$ is between
2 and a maximum value $q_{\rm max}$.  We show in the Appendix that the
uncertainty  in  $q$ is  an  intrinsic  property  of the  EPS  theory;
different algorithms  for constructing merger-trees  may correspond to
different $q$  within the allowed  range.  The maximum  value, $q_{\rm
  max}$, depends slightly on  cosmology and mass. For example, $q_{\rm
  max}$  is  between  $2.1$   and  $2.3$  for  flat  cosmogonies  with
$\Omega_m$  between 0.1  and 0.9  and halo  masses between  $10^8$ and
$10^{15}\hmsun$.

Solving the differential equation for $\bmbig$ we come up with a useful
fitting formula:
\begin{equation}
\label{eq:mbig}
\bmbig(z) = \frac{\Omega_m}{\Gamma^3}F_q^{-1} \biggl[
\frac{g(32\Gamma)}{\sigma_8}(\omega(z)-\omega_0) + F_q\bigl(
\frac{\Gamma^3}{\Omega_m} M_0 \bigr) \biggr] \,.
\end{equation}
Here $g$  and $F_q$  are analytic fitting  functions motivated  by the
shape of  the power-spectrum (see  Appendix for their  definition, and
range  of accuracy).  For the  $\Lambda$CDM concordance  cosmology, we
find  that the  standard  algorithm of  SK99  for constructing  random
merger trees yields an $\bmbig(z)$  which is well fitted by \equ{mbig}
with  $q=2.2$.  We  therefore  adopt this  value  below.  Varying  $q$
between  2  and  2.3 (the  maximum  range  allowed)  gives rise  to  a
relatively small change in  $\bmbig$; near $\bmbig=0.5M_0$ this change
is $\sim8\%$.

\begin{figure}
\centerline{ \hbox{ \epsfig{file=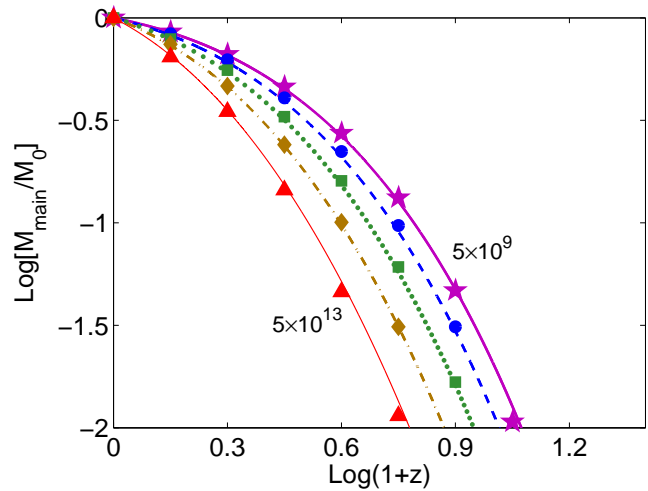,width=9cm} } }
\caption{Growth of the main progenitor for haloes of different
  present-day masses. The  mass $\bmbig(z)$ is the average  at a fixed
  redshift $z$. The curves are  normalized to match at $z=0$. The halo
  masses,   from  top   to   bottom,  range   from  $5\times10^9$   to
  $5\times10^{13}\hmsun$ equally spaced in the log.  The symbols refer
  to  the  averages  over  Monte-Carlo  merger trees  and  the  curves
  represent our analytic results.  The upsizing of the main progenitor
  is obvious.} \label{figs:mbig}
\end{figure}

\begin{figure}
\centerline{ \hbox{ \epsfig{file=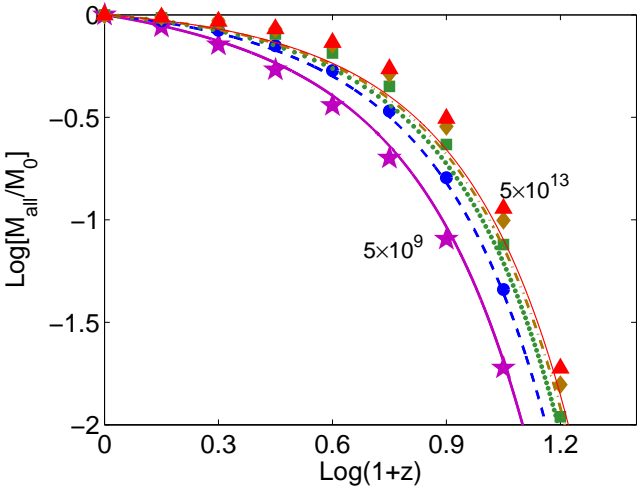,width=9cm} } }
\caption{Growth of the total mass in all the progenitors, $\bmall(z)$,
  for haloes of different  present-day masses.  The minimum progenitor
  mass is $\sim 10^9\msun$, specified in equation~(\ref{eq:mmin}) as a
  function of redshift. The masses, curves and symbols are the same as
  in Fig.~\ref{figs:mbig}. A downsizing behaviour is clearly seen.  It
  is more pronounced at small masses which are closer to $\mmin$.}
\label{figs:mall}
\end{figure}

Fig.~\ref{figs:mbig}   shows   $\bmbig(z)/M_0$,   the  average,   main
progenitor  history   for  haloes  of  different   masses  today,  all
normalized to today's mass.  The figure compares our analytic estimate
based  on equation~(\ref{eq:mbig})  with the  averages  over histories
computed  from  Monte-Carlo  merger  tree  realizations  described  in
\se{merger_trees}.  We  see that the analytic  estimate reproduces the
results from the  realizations quite well, although there  is a slight
mismatch at high  $z$. This difference may either  reflect the allowed
intrinsic uncertainty  within the  EPS formalism or  it may be  due to
other inaccuracies in the SK99 algorithm used to construct the trees.

\subsection{Archaeological Upsizing}
\label{sec:upsizing}

We  see  in  Fig.~\ref{figs:mbig} that the average growth curve of the
main progenitor  is shifted toward later times in more massive haloes,
implying  the  opposite of downsizing, termed here as upsizing. One
way to quantify  the  downsizing  behaviour is via the quantity
\begin{equation}
\label{eq:ds_strength}
\dmain(z\,\vert\, z_0,M_0) \equiv
\frac{\dd}{\dd M_0} \left[ \frac{\bmbig(z)}{M_0} \right] \,.
\end{equation}
Positive   values  of  $\dmain$   mark  an   archeological  downsizing
behaviour, negative  values refer to  upsizing, and $\vert\dmain\vert$
measures   the   strength  of   the   effect.    As   is  clear   from
Fig.~\ref{figs:mbig}, $\dmain(z) < 0$  at all $z$, indicating that the
main progenitor histories of dark matter haloes reveal upsizing.

Is this upsizing a generic  feature of $\bmbig(z)$? To answer this, we
write the average main progenitor mass of a halo of mass $M_0$ a small
time step $\dW$ ago as
\begin{equation}
\label{asmblhis2}
\frac{\bmbig(\dW)}{M_0} = \int_{0}^{S_q-S_0}f(\dS,\dW)\dd\dS \,,
\end{equation}
(see Appendix).  Differentiating with  respect to $M_0$  while keeping
$\dW$ fixed yields the ADS strength
\begin{eqnarray}
\label{eq:big_derivative}
\lefteqn{ \dmain(z\,\vert\,z_0,M_0) = } \\
\nonumber & & f(S_q-S_0,\dW) \left[ \frac{1}{q}\frac{\dd S}{\dd
M}(M_0/q) - \frac{\dd S}{\dd M}(M_0) \right] \,.
\end{eqnarray}
Whether this is negative or not  depends on the shape of $S(M)$. For a
self-similar  power spectrum,  $S \propto  M^{-\alpha}$, we  have that
$\dmain < 0$ as long as $\alpha > 0$. We  have also verified
numerically that  $\dmain<0$ for  the standard $\Lambda$CDM power
spectrum at  all masses. While the above expression for $\dmain$  is
valid only for small  $\dW$, its sign is  the same at all $z$. A more
accurate expression for $\dmain$ at any  $z$ can be obtained by
differentiating equation~(\ref{eq:mbig}) above.

\subsection{Assembly Time of the Main Progenitor}
\label{sec:assemblytime}

Following   numerous   other   studies  (see  \se{intro}),  we  define
the  assembly  redshift  $\zbig$  of  a  halo  of  mass  $M_0$ at time
$\omega_0$      according      to      $\mbig(\zbig)=M_0/2$.     Using
equation~(\ref{eq:mbig}) we obtain
\begin{eqnarray}
\label{eq:assred}
\lefteqn{ \bwbig \equiv \omega(\bzbig) = } \\
\nonumber & & \omega_0 + \frac{\sigma_8}{g(32\Gamma)}\left[
F_q\left(\frac{\Gamma^3}{\Omega_m} \frac{M_0}{2}\right) -
F_q\left(\frac{\Gamma^3}{\Omega_m}M_0\right) \right] \,.
\end{eqnarray}
In the case of  scale-free initial conditions, were the power-spectrum
is  a pure power  law, $P(k)  \propto k^{n}$,  we have  that $S\propto
M^{-\alpha}$  with $\alpha  = (n+3)/3$.  In this  case  the expression
simplifies to
\begin{equation}
\label{eq:wmain} \bwbig=\omega_0 + {\sqrt{2\pi (q^{\alpha} - 1)}
\over \alpha} \left( \sqrt{S_q}-\sqrt{S_0}\right) \,.
\end{equation}
We  note  that  \citet{Lacey93}  computed a related expression for the
average assembly redshift of the main progenitor (see the Appendix for
more details).

\begin{figure}
\centerline{ \hbox{ \epsfig{file=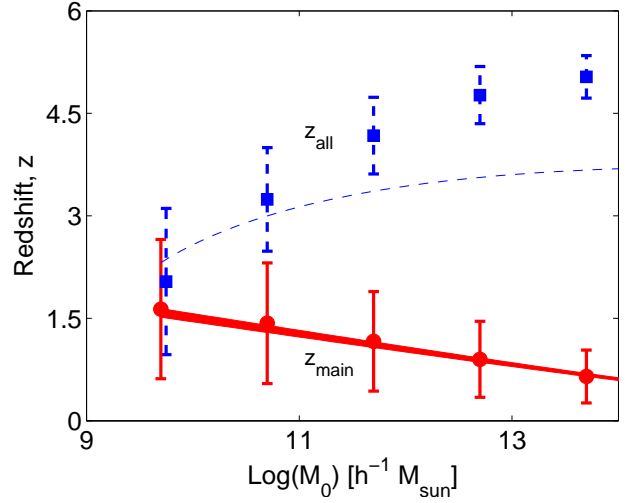,width=9cm} } }
\caption{Formation redshifts, when the mass was one half of today's
  mass, for  the main progenitor ($\bzbig$, solid  lines, circles) and
  for  all the  progenitors ($\bzall$,  dashed lines,  squares) versus
  today's halo mass.  The symbols  and error-bars refer to an ensemble
  of  random EPS merger-trees.   The thickness  of the  $\bzbig$ curve
  refers to  the allowed range obtained  by varying $q$  between 2 and
  $q_{\rm max}$ in equation~(\ref{eq:assred}).  The dashed line is the
  theoretical prediction (\ref{eq:zfall}) for $\bzall$.  $\zbig$ shows
  upsizing while $\zall$ shows downsizing.}
\label{figs:z_average}
\end{figure}

Fig.~\ref{figs:z_average} shows  the average assembly  redshift of the
main progenitor, $\bzbig$, as a function of the present-day halo mass,
for the $\Lambda$CDM concordance cosmology. The thickness of the curve
corresponds to  the allowed range  of intrinsic uncertainty in  $q$ in
equation~(\ref{eq:assred}),  as  computed in  the  Appendix.  The  ADS
strength, $\dmain$, associated with the slope of $\bzbig(M)$, does not
change significantly with halo  mass.  The theoretical EPS curve shows
an excellent agreement with the  $\bzbig$ obtained from an ensemble of
random EPS merger histories (circles  with error bars). The error bars
correspond to  the standard deviation  in $\zbig$ over  the individual
merger-trees.

Fig.~\ref{figs:z_average} demonstrates again the upsizing behaviour of
the main progenitor. This has been one of the reasons for interpreting
the observed downsizing as ``anti-hierarchical".
\begin{figure}
 \centerline{ \hbox{ \epsfig{file=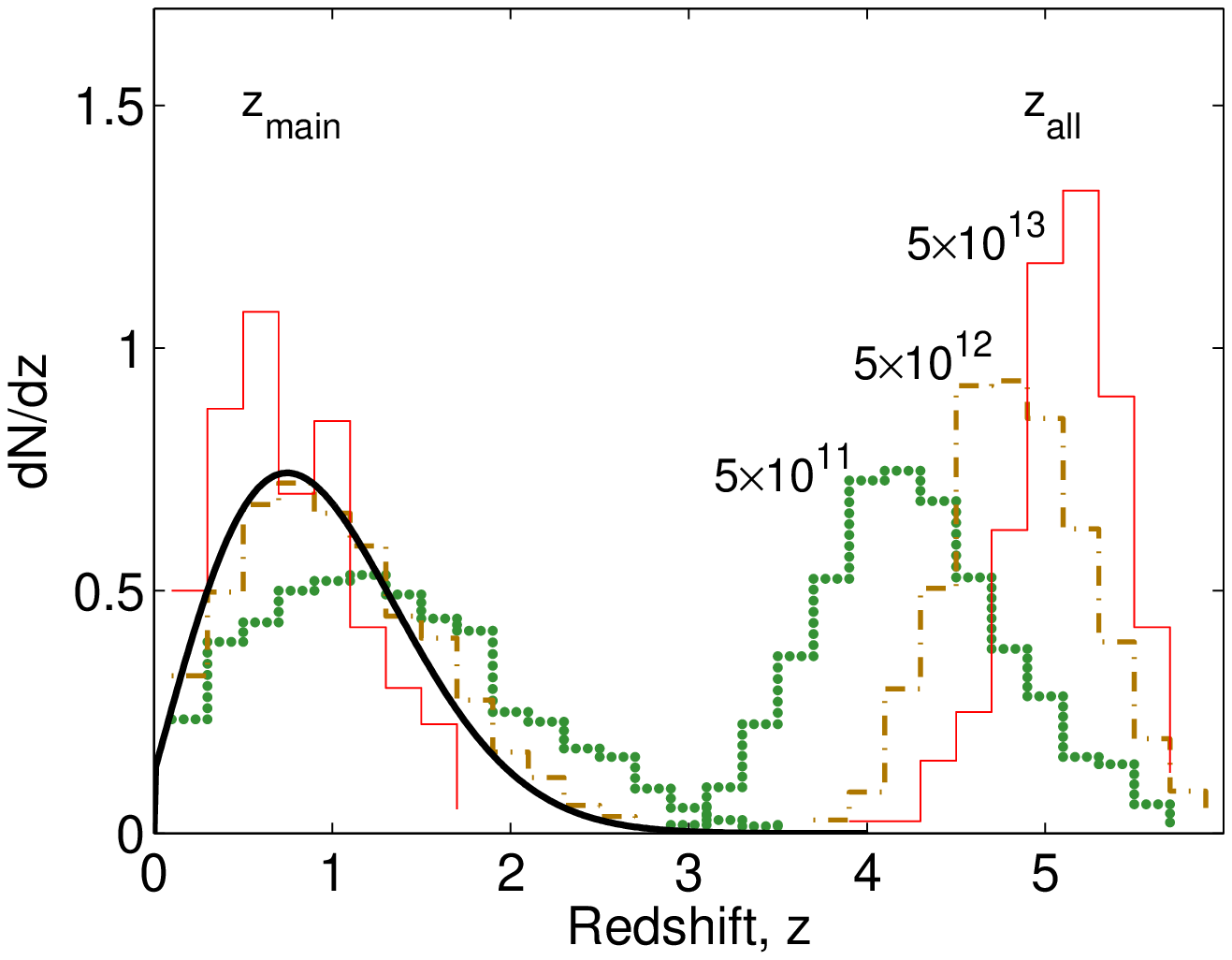,width=9cm} } }
  \caption{The distribution of $\zbig$ and $\zall$ for different halo
    masses.   Halo masses  are $5\times10^{11}$,  $5\times10^{12}$ and
    $5\times10^{13}\hmsun$  (dotted, dashed-dotted  and solid  lines).
    The  solid  thick  line  is  the theoretical  prediction  for  the
    distribution of $\zbig$, equation~(\ref{eq:Q}), for a halo mass of
    $5\times10^{12}  \hmsun$.  The  minimum progenitor  mass  is $\sim
    10^9\msun$, specified in equation~(\ref{eq:mmin}) as a function of
    redshift.}
  \label{figs:z_form_scatter}
\end{figure}

The distribution  of $\zbig$  in our ensemble  of EPS merger  trees is
plotted in Fig.~\ref{figs:z_form_scatter}  for three different masses.
One of them is compared to the theoretical prediction by LC93,
\begin{equation}
\label{eq:Q} Q(z) = -\frac{\dd}{\dd z} \int_{S_0}^{S_2}
\frac{M_0}{M}f(S-S_0,\omega(z)-\omega_0)\dd S \,,
\end{equation}
where $f$  is defined  in equation~(\ref{eq:probS}).  As  discussed in
the  Appendix, the  theoretical  distribution agrees  with the  random
realizations at low $z$, and any deviations are due to the limitations
of the SK99 algorithm used.

\section{Growth of All the Progenitors}
\label{sec:mall}

$\mbig(z)$ defined above only describes the mass growth history of the
main  trunk  of  the  full  merger tree. It is unlikely, however, that
this  is  an  honest  estimator of the star formation histories of the
associated  galaxies. After all, star formation can occur in {\it all}
progenitors  that  obey  the necessary physical conditions, and is not
restricted  to  the  {\it  main}  progenitor.  Since gas needs to cool
before  it  can  form stars, and since the cooling time is primarily a
function  of  halo  mass  and  redshift, we assume that star formation
occurs  in haloes with a mass above a threshold mass, $\mmin(z)$. This
prompts us to define the formation history $\mall(z)$ of a present-day
dark matter halo as the sum of the masses of all progenitors that obey
this  condition.  Supporting evidence for the possible success of such
a  model  comes  from the finding that the integral of $\mall(z)$ over
the  entire  present-day halo mass function provides a useful backbone
for  understanding  the  observed,  universal  star  formation history
\citep{Hernquist03}.

\subsection{Analytical Derivation}
\label{sec:analytic_mall}

The  construction  of  $\mall(z)$  for  individual  dark matter haloes
requires  a  full  merger  tree,  with  a mass resolution that exceeds
$\mmin(z)$.  However,  the  formation history of a halo of mass $M_0$,
averaged  over many merger trees per each redshift $z$, can be derived
straightforwardly   from  the  EPS  formalism.  It  should  equal  the
integral over the progenitor mass function in the range $M = \mmin(z)$
to $M_0$:
\begin{equation}
\label{eq:mall} \bmall(z) = M_0 \int_{\mmin(z)}^{M_0} P(S,\omega \vert
S_0,\omega_0) \abs{\frac{\dd S}{\dd M}} \dd M \ \,,
\end{equation}
where    $P(S,\omega    \vert    S_0,\omega_0)$    is    defined    in
equation~(\ref{eq:probSS}). Performing the integral we obtain
\begin{equation}
\label{eq:mall_erf}
\frac{\bmall(z)}{M_0} = 1 - {\rm erf}
\left(\frac{\omega(z)-\omega_0}{\sqrt{2 \smin(z)-2 S_0}}\right)\ ,
\end{equation}
where $\smin(z) \equiv S[\mmin(z)]$. Note that $\bmall(z)/M_0$ depends
on $M_0$ through $S_0$, so that equation~(\ref{eq:mall_erf}) cannot be
written in an explicit form.

We see that for a given  cosmology, the average formation history of a
halo of mass $M_0$ is  completely specified by $\mmin(z)$.  As a first
attempt we associate $\mmin$ with  the halo mass that corresponds to a
virial temperature of  $T_{\rm vir} = 10^4 \K$,  the temperature above
which atomic  gas is able to  cool and subsequently form  stars. For a
completely ionized, primordial gas this yields
\begin{equation}
\label{eq:mmin}
\mmin(z) =  1.52 \times 10^9 h^{-1} \msun
\left({\Delta_{\rm vir} \over 101}\right)^{-1/2} \left({H(z) \over
H_0}\right)^{-1}\,,
\end{equation}
where $\Delta_{\rm vir}(z)$ is the average overdensity of a virialized
halo at redshift $z$ relative to the critical density at that redshift
\citep{Bryan98},   and   $H(z)$   is   the  Hubble  parameter.  Unless
specifically  stated  otherwise, we use this minimum threshold mass in
what follows.

The lines  in Fig.~\ref{figs:mall}  show $\bmall(z)$ for  several halo
masses based  on equation~(\ref{eq:mall_erf}).   We now see  that when
``formation"  is  defined by  $\bmall(z)$  we  obtain  ADS, with  more
massive   haloes    forming   {\it   earlier}.     Also   plotted   in
Fig.~\ref{figs:mall} are  the results  of the merger-trees  (symbols).
We see that  while there is a fair,  qualitative agreement between the
histories extracted  from the  Monte-Carlo realizations and  the exact
EPS predictions,  the level  of agreement becomes  progressively worse
for  more massive  haloes  (relative to  $\mmin$).   The merger  trees
predict an earlier formation time than what follows directly from EPS.
A  similar behaviour  has been  noticed by  SK99 (their  Fig.~7), when
comparing the empirical total mass contained in haloes above a minimum
mass  to  the theoretical  value.   These  deviations  arise from  the
approximations made in the  algorithm for constructing the Monte-Carlo
merger trees (see discussion in \se{merger_trees}).

\begin{figure}
 \centerline{ \hbox{ \epsfig{file=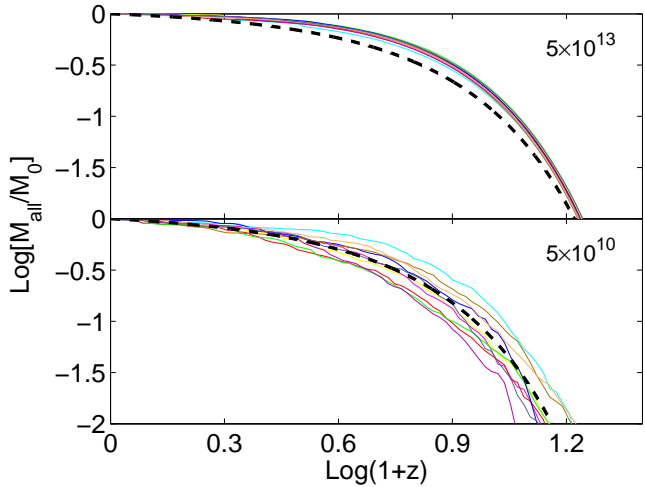,width=9cm} } }
\caption{Individual realizations of all-progenitor histories (thin solid
  curves)  compared  to  the   average  at  fixed  $z$  as  calculated
  analytically (thick dashed curve).  The curves are all normalized to
  $M_0$   at   $z=0$.    The   upper   and  lower   panels   are   for
  $M_0=5\times10^{13}$   and   $5\times10^{10}\hmsun$  respectively.
  $\mmin$ is specified in equation~(\ref{eq:mmin}). }
  \label{figs:mall1013}
\end{figure}

Figure~\ref{figs:mall1013}   shows   the   $\mall(z)$   histories   of
individual haloes, obtained  from Monte-Carlo merger tree realizations
(thin lines),  compared to the average  formation history $\bmall(z)$,
calculated from eq.~\ref{eq:mall_erf} (thick dashed line). The scatter
is higher for  lower mass haloes. This can be  crudely understood as a
Poisson noise associated with $N_{\rm all}$, the number of progenitors
above  $\mmin$ at  every given  redshift. For  the massive  halo, $M_0
=5\times10^{13}\hmsun$,  $N_{\rm all}$  is indeed  quite large  at all
redshifts, leading to a small scatter. For the less massive halo, $M_0
= 5\times10^{10}\hmsun$,  we have  $N_{\rm all}<20$ at  all redshifts,
which results in a larger scatter.

\subsection{Archaeological Downsizing}
\label{sec:DS}

The all-progenitor  histories $\mall(z)$  depend on the  definition of
the  threshold mass  $\mmin(z)$.   Here we  investigate the  necessary
conditions  for these  threshold masses  in order  for  $\bmall(z)$ to
reveal ADS.  Similar to what  was done in \se{upsizing}, we define the
``downsizing  strength'', $\dall$, as  the derivative  of $\bmall/M_0$
with respect  to $M_0$.   In order to  study the $M_0$  dependence, we
rewrite equation~(\ref{eq:mall}) using different variables,
\begin{equation}
\label{eq:mall2}
\frac{\bmall(\omega)}{M_0} =
\int_{0}^{\smin-S_0}f(\dS,\omega-\omega_0)\dd\dS \ ,
\end{equation}
where the function $f$  is defined in equation~(\ref{eq:probS}).  This
enables us  to differentiate $\mall/M_0$ with respect  to $M_0$, while
keeping $\omega$ fixed, which yields
\begin{eqnarray}
\label{eq:ds_derivative} \dall(\omega\,\vert\,\omega_0,M_0) = -
f(S_{\rm min}-S_0,\omega-\omega_0) \frac{\dd S}{\dd M}(M_0)>0\,.
\end{eqnarray}
Since $f$  is a probability  function, and $\dd  S/\dd M < 0$  for all
$M$, we  have that $\dall$ is  always positive. This  implies that ADS
occurs for any choice of  the threshold masses $\mmin(z)$, and for any
cosmological power spectrum of fluctuations. The only assumptions used
are  (i) that the  threshold is  global, i.e.,  that $\mmin$  does not
depend  on   the  specific  halo   mass  $M_0$,  and  (ii)   that  the
excursion-set trajectories  are Markovian, which allows  the change of
variables leading  from equation~(\ref{eq:mall}) to  (\ref{eq:mall2}).
Note that the downsizing aspect  of $\bmall(z)$ does not depend on the
actual  shape  of   $f$,  which  implies  that  ADS   will  occur  for
non-Gaussian density fluctuation fields as well.

The  opposite  effect  of  upsizing  could  in  principle occur if the
Markovian  assumption of the EPS random walks breaks down, so that the
mass-weighted probability distribution $P(S,\omega\vert S_0,\omega_0)$
depends  on  $S_0$  rather  than  being a function of $S-S_0$ only. An
additional  requirement  in  this  case  is that the probability has a
higher contribution from the low-mass end for larger $M_0$. Therefore,
the  Markovian  nature  of the random walks is a sufficient, but not a
necessary condition for ADS to occur.

\subsection{Formation Time of All progenitors}

Following the definition of assembly redshift, we define the formation
redshift  of  dark  matter  haloes,  $\zall$, by $\mall(\zall)=M_0/2$.
Using equation~(\ref{eq:mall_erf}) we obtain
\begin{equation}
\label{eq:zfall}
 \bwall \equiv \omega(\bzall) = \omega_0 + \beta \sqrt{\smin-S_0} \,,
\end{equation}
where   $\beta=\sqrt{2}/{\rm   erf}(1/2)   \simeq  0.6745$  \citep[see
also][]{Bower91}.  The dashed curve in Fig.~\ref{figs:z_average} shows
$\zall$    as    a    function    of    halo   mass   computed   using
equation~(\ref{eq:zfall}).    The    solid   square   with   errorbars
represent  the  average  and scatter as obtained from a large ensemble
of  EPS  merger  trees. Note that these deviate significantly from the
direct  theoretical EPS prediction, especially at large $M_0$. This is
in  stark  contrast  to  the  case  of  $\zbig(M_0)$, where the merger
tree    results    agree    well    with    the   direct   theoretical
predictions.  This  suggests that the discrepancy in $\zall(M_0)$ must
originate  in  the  statistics  of  smaller  progenitors  with  masses
$<M_0/2$.  As shown by SK99, the $N$-branch tree method with accretion
used   for   the   construction  of  the  EPS  merger  trees  slightly
overpredicts  the  number  of  small  progenitors  at  high redshifts.
Fig.~\ref{figs:z_average}  shows  that  this  can  have  a significant
impact  on  $\zall$;  consequently,  semi-analytical models for galaxy
formation  that  are  based  on  such  EPS merger trees might actually
overestimate the star formation rates at high redshifts.

For  haloes  with  $M_0 \gg  \mmin$  we  have  that $\zall  >  \zbig$:
typically  the   progenitors  of  a  massive  halo   will  have  grown
sufficiently massive to allow for star formation much before the final
halo has assembled half its present-day mass into a single halo.  Note
that $\zall -  \zbig$ decreases with decreasing halo  mass.  When $M_0
\simeq  2\mmin$ we  have that  $\zall =  \zbig$, by  definition, while
$\zall < \zbig$  for haloes with $\mmin < M_0  < 2\mmin$. Finally, for
haloes  with $M_0  <  \mmin$  the formation  redshift  $\zall$ is  not
defined. This  systematic increase of $\zall -  \zbig$ with increasing
halo mass  may have interesting implications for  galaxy formation, as
it provides a very natural  means to break the self-similarity between
haloes of different masses, and their associated galaxies.

Although  dynamical friction may  delay the  merging of  galaxies with
respect to the epoch at which their host haloes merged, to first order
we may associate $\zbig$ with  the redshift below which the haloes and
their associated  galaxies no  longer experience major  mergers (i.e.,
below $\zbig$  the main progenitor  never merges with another  halo of
similar mass).  In massive haloes, with $M_0 \gg \mmin$ we expect that
the majority of  the stars have already formed  much before these last
major  mergers,  and  this  majority  of  the  stars  will  thus  have
experienced  one  or  more   major  mergers  since  their  formation.
Consequently, the majority of the stars are most likely to reside in a
spheroidal component, and the  galaxy is an early-type with relatively
old stars. Contrary, in low mass haloes, most of the progenitor haloes
that are  being accreted by  the main progenitor  at $z <  \zbig$ will
have masses $M < \mmin$, and  will thus not have formed stars. The gas
associated with  these progenitors can  only start to form  stars once
they become  part of  the main progenitor:  star formation  and galaxy
assembly  occur  virtually hand-in-hand,  with  the  stars being  born
in-situ in  what is to  become the final  galaxy at $z=0$.   Since the
system has not  undergone a major merger since  roughly half the stars
formed, the system is likely to resemble a disk galaxy.

Although  this is  clearly severly  oversimplified, it  is interesting
that some of the most pronounced scaling relations of galaxies, namely
the relations  between halo mass, stellar age,  and galaxy morphology,
may well  have their direct origin  in the backbone  of halo formation
histories  combined  with  a  simple  halo  mass  threshold  for  star
formation.

Finally,  Fig.~\ref{figs:z_form_scatter}  shows  the  distribution  of
$\zall$  for haloes  of different  masses,  as obtained  from our  EPS
merger trees.   Note that the scatter  in $\zall$ is  smaller for more
massive haloes,  as expected from the Poisson  statistics discussed in
\S\ref{sec:analytic_mall}

\subsection{Comparison with $N$-body Simulations}
\label{sec:mall_Nbody}

While  the  merger  trees analyzed  thus  far  are  based on  the  EPS
formalism,   one   can  alternatively   extract   merger  trees   from
cosmological $N$-body simulations.  Here the gravitational dynamics is
more accurate, limited only by numerical resolution effects.  However,
it  should be  kept  in  mind that  the  identification of  virialized
haloes, and especially connecting them to construct merger trees, is a
non-trivial enterprise involving several significant uncertainties.

\begin{figure}
 \centerline{ \hbox{ \epsfig{file=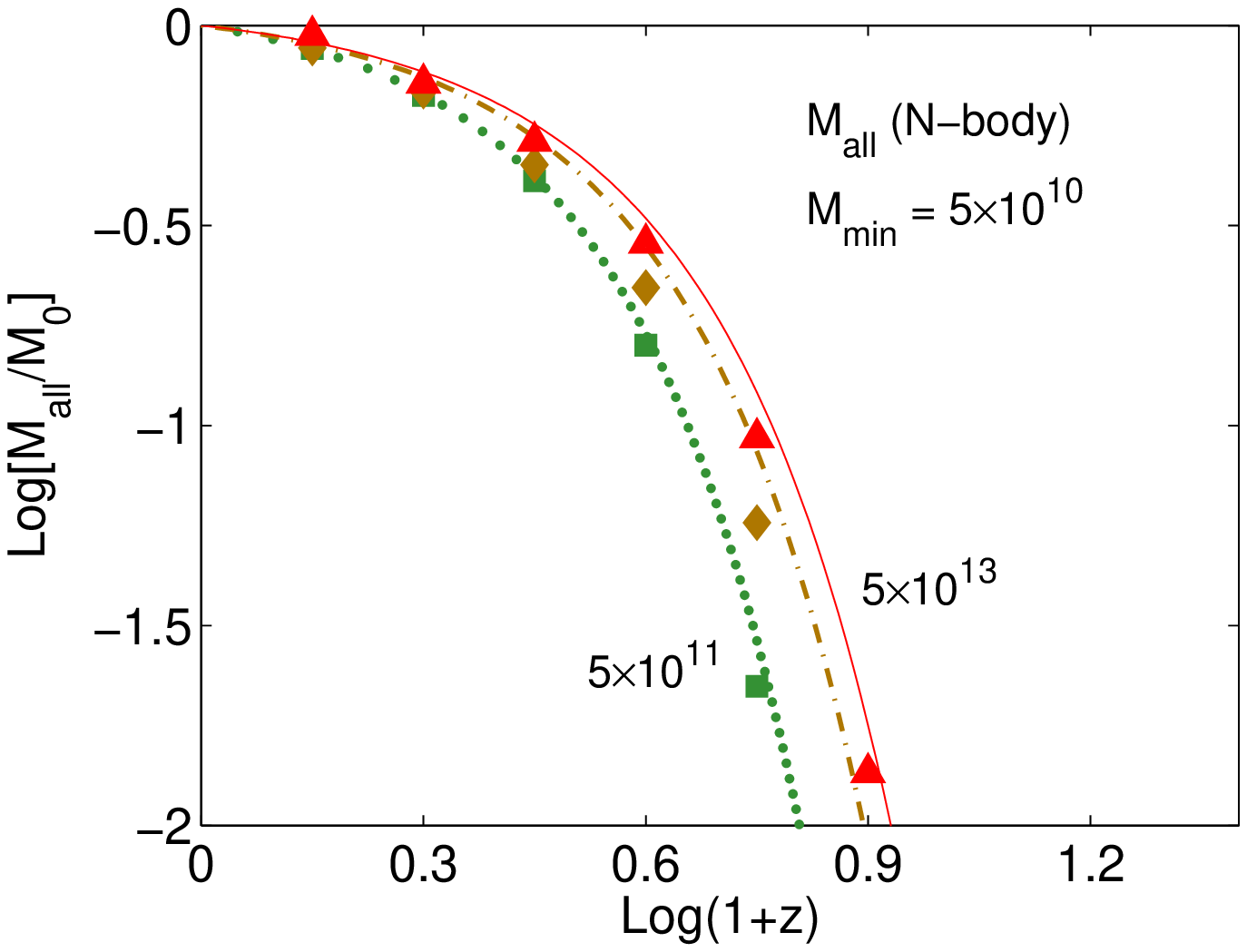,width=9cm} } }
 \centerline{ \hbox{ \epsfig{file=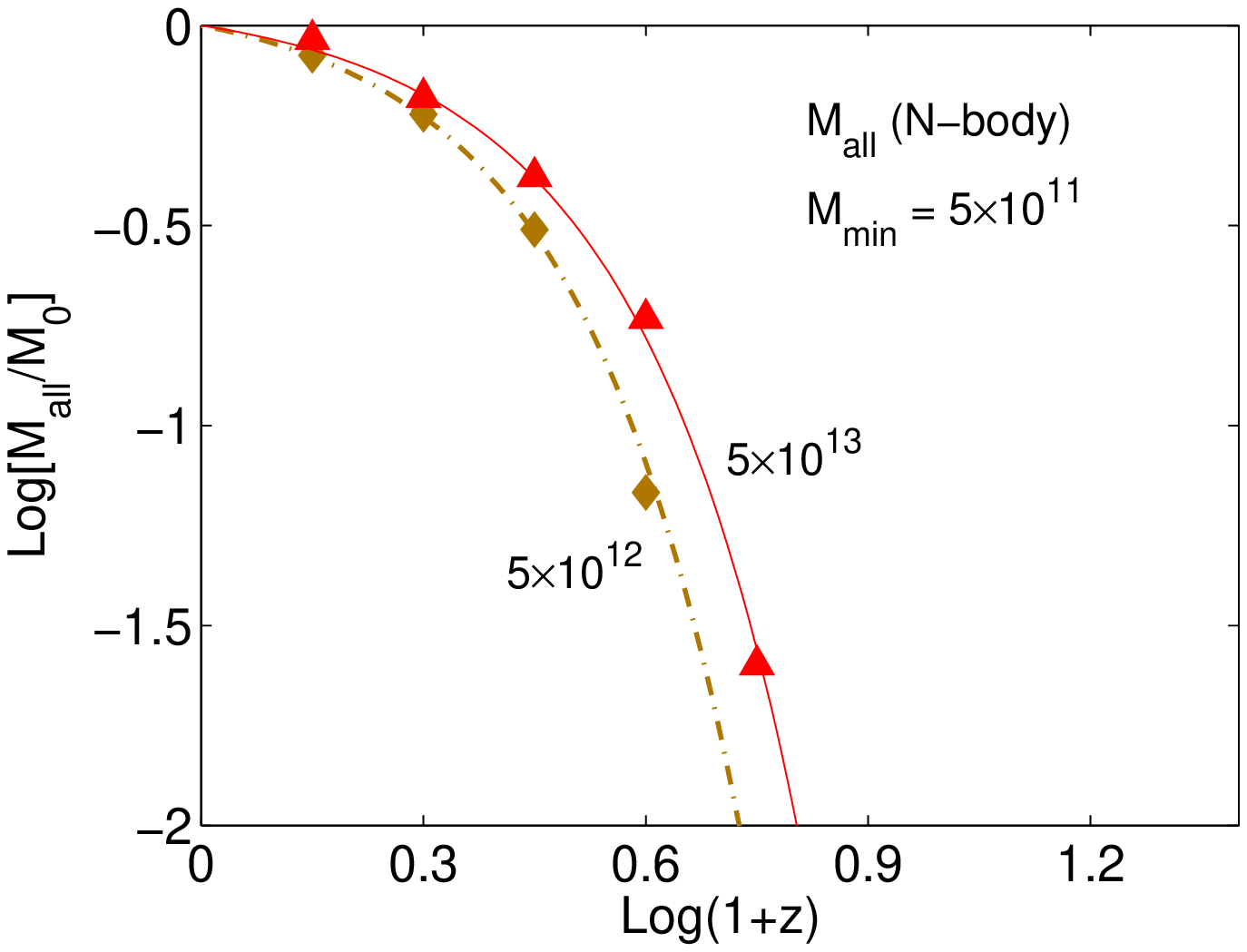,width=9cm} } }
\caption{All-progenitor histories drawn from $N$-body simulations
  (symbols)  compared to  the EPS  predictions (curves).   The imposed
  minimum mass is  $\mmin=5\times10^{10}$ and $5\times10^{11}\msun$ in
  the top  and bottom panels respectively.  The mass bins  in log mass
  are $[11.62,11.78],  \; [12.48,13.00], \;  [13.48,14.00]$ where mass
  units are $\hmsun$. The number of haloes within each bin is 479, 205
  and  23 respectively.  The EPS  theoretical curves  corresponding to
  each mass bin are averages over the same distribution of masses.}
  \label{figs:mall_Nbody}
\end{figure}

We compute $\bmall(z)$ from merger trees extracted from a $\Lambda$CDM
cosmological  $N$-body simulation  kindly provided  by Risa  Wechsler.
The simulation  followed the trajectories of $256^3$  cold dark matter
particles within a cubic, periodic box of comoving size $60\hmpc$ from
redshift   $z=40$   to   the    present.    The   particle   mass   is
$1.1\times10^{9}\hmsun$,  and the  minimum halo  mass dictated  by the
resolution  is  $2.2\times10^{10}\hmsun$  \citep[see][for
details]{Wechsler02}. For the construction  of  $\bmall(z)$  we impose
$\mmin$ values of $5\times10^{10}$  and $5\times10^{11}\hmsun$, and we
compare the resulting, average formation  histories to those computed
from the EPS formalism using  the same  threshold masses. The results
are  shown in Fig.~\ref{figs:mall_Nbody},  where symbols correspond to
the formation  histories extracted  from the  $N$-body simulations,
while the lines  show the direct, theoretical predictions based on the
EPS formalism (equation \ref{eq:mall_erf}).  Overall the agreement is
very satisfactory,  although  the $N$-body  simulations predict  a
somewhat later formation  when $\mmin  \ll  M_0$. Note  that the  EPS
\emph{merger trees} yield formation  times that are {\it earlier} with
respect to  the analytical formula  (Fig.~\ref{figs:mall}).  Thus, the
difference between $N$-body simulations and EPS merger trees is larger
than   the   difference   between   the   $N$-body   simulations   and
equation~(\ref{eq:mall_erf}).    Despite   these  discrepancies,   the
$N$-body results  clearly confirm the  EPS prediction that  $\mall$ of
more massive haloes grows earlier.  We therefore conclude that the ADS
aspect of $\mall$ is not an  artifact of the EPS approximation, but is
a generic property of DM merger trees.

\subsection{The Correlation between Formation Time and Assembly Time}

Since $\zall$ increases with increasing halo mass (ADS), while $\zbig$
decreases  (`upsizing'),   we  have  that  $\zall$   and  $\zbig$  are
anti-correlated when considering haloes  of different masses. But what
about the relation  between $\zall$ and $\zbig$ for  haloes of a fixed
mass?

\begin{figure}
 \centerline{ \hbox{ \epsfig{file=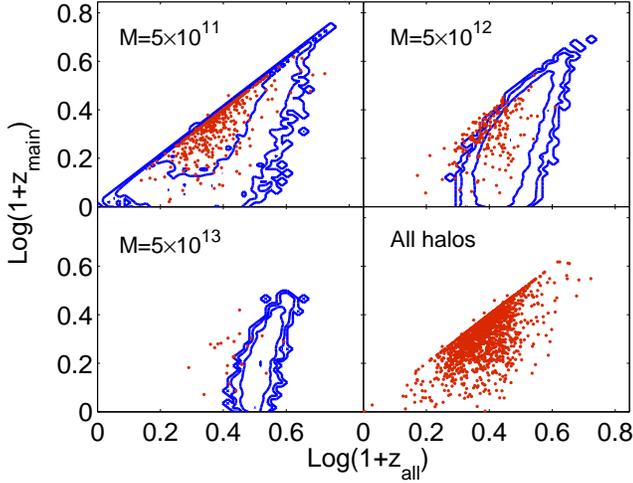,width=9cm} } }
\caption{Correlations between $\zbig$ and $\zall$ for
  random merger trees with the same final mass.  The contours, equally
  spaced  in the log,  refer to  the joint  distribution from  the EPS
  random merger  trees.  The  minimum mass is  $\mmin=5\times10^{10}$.
  The points are  from $N$-body merger-trees, with the  same mass bins
  as in Fig.~\ref{figs:mall_Nbody}. }
  \label{figs:form_asmbl_corr}
\end{figure}

Fig.~\ref{figs:form_asmbl_corr} shows  the correlation between $\zall$
and $\zbig$  for haloes of given masses,  with $\mmin=5\times10^{10}$.
Results are shown for both  the numerical simulations (solid dots) and
for EPS  merger trees (contours).  For  a $5\times10^{11}\hmsun$ halo,
the number  of progenitors is small,  and the full merger  tree is not
much more than the main trunk.  As a result, the values of $\zall$ and
$\zbig$  are not  very  different  and they  exhibit  a rather  strong
correlation. When the  mass gets larger, the scatter  in $\zall$ tends
to zero while the scatter in $\zbig$ remains large.  Consequently, the
correlation strength  between $\zbig$ and  $\zall$ at fixed  halo mass
vanishes at large $M_0$.

This has  important implications. Using a  large numerical simulation,
\citet{Gao05} and  \citet{Harker06} have found  a positive correlation
between  $\zbig$  and the  environment  density:  i.e.,  haloes in  an
overdense region assemble  earlier than haloes of the  same mass in an
underdense  region. If  galaxy  properties, such  as  stellar age,  is
correlated with $\zbig$,  this means that haloes of  a given mass host
galaxies  with different  properties, depending  on their  large scale
environment. The results shown here  suggest that this may be the case
for relatively  low mass haloes  with $M_0 \simeq \mmin$,  since these
systems reveal a positive correlation between $\zbig$ and $\zall$.  In
more  massive   haloes,  however,  with  $M_0  \gg   \mmin$,  no  such
correlation  is  present,  suggesting  that  the  correlation  between
$\zbig$ and environment will  not create a similar correlation between
stellar age and  environment.

The  positive correlation between  $\zbig$ and  $\zall$ at  fixed mass
arises  from their  dependence on  the merger  tree properties  at low
redshifts.  For example, assume that  the merger tree for some halo is
such that the mass  at $z=0.5$ is the same as at  $z=0$. In this case,
we can use the analytical expressions for $\zall$ and $\zbig$ starting
at  $z_0=0.5$ ,  and not  at  $z_0=0$. The  corresponding $\zbig$  and
$\zall$ will refer  to $z_0=0.5$, so both will be  delayed by the same
amount  of time, thus  establishing a  positive correlation.

When we increase the halo mass, the ratio $\bmbig(z)/\bmall(z)$
decreases (this is true for any specific redshift $z$). This implies
that the fraction of mass incorporated in the main trunk is smaller,
and as a result, there is more mass left in progenitors that belong to
other branches. The scatter in $\bmall$ comes from all the tree
branches, where each branch contribute its own random behaviour. When
$\bmbig(z)$ is small with respect to $\bmall(z)$ most of the
contribution to the scatter in $\bmall$ comes from branches other than
the main. This explains why the correlation between $\zall$ and
$\zbig$ gets poorer for high halo mass.

\section{Halo Formation Rates}
\label{sec:formation_rate}

We define  the halo formation rate as  the rate of change  in $\mall$.
Using equation~(\ref{eq:mall_erf}), this can be written as
\begin{eqnarray}
\label{eq:mall_rate}
\lefteqn{ R(\omega \,\vert\, \omega_0,M_0 ) \equiv
\frac{\dd}{\dd t} \left[ \frac{\mall(\omega)}{M_0} \right] =} \\
\nonumber & & - \sqrt{\frac{2}{\pi}} \frac{1}{\sqrt{\smin-S_0}}
\exp \left[ - \frac{(\omega-\omega_0)^2}{2\smin-2S_0} \right]
\frac{\dd\omega}{\dd t} \,.
\end{eqnarray}
If  we make the  naive assumption  that all  the baryonic  mass inside
haloes with  $M > \mmin$  forms stars instantaneously, than  this rate
reflects  the  star  formation   history  of  galaxies  that  at  time
$\omega_0$ are located in a  halo of mass $M_0$: these formation rates
basically reflect the maximum possible star-formation efficiency.

\begin{figure}
 \centerline{ \hbox{ \epsfig{file=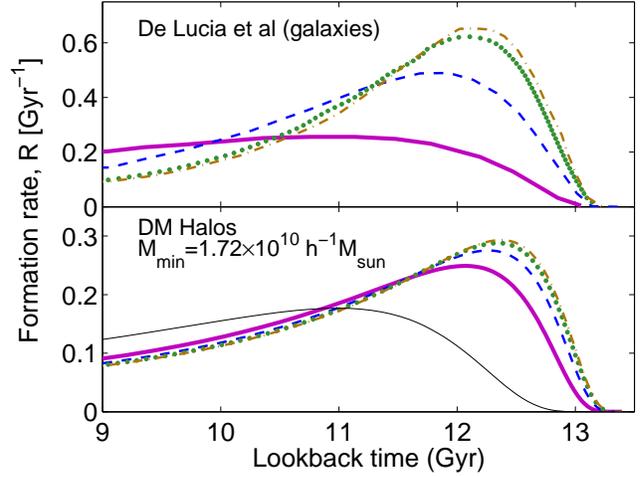,width=9cm} } }
  \caption{Simulated star formation rate versus EPS halo formation rate $R$
    for  different halo masses.   Top panel:  mean SSFR  of elliptical
    galaxies taken from the  semi-analytic model of \citet{DeLucia06}.
    Galaxies are  binned by their  halo mass at $z=0$.   Bottom panel:
    maximum  SFR as  implied by  our simplified  model, namely  $R$ of
    equation~(\ref{eq:mall_rate}), for different halo masses at $z=0$.
    The curves  in the two panels  refer to halo  masses of $10^{12}$,
    $10^{13}$, $10^{14}$, and $10^{15}$ $\msun$(solid, dashed, dotted,
    dashed-dotted      lines      respectively).       $\mmin$      is
    $1.72\times10^{10}\hmsun$, the minimum halo mass in De Lucia \etal
    (2006).  In  the lower  panel we add  a curve  for a halo  mass of
    $10^{12}\;\msun$ with $\mmin$  set to $10^{11}\;\msun$ (thin solid
    line), as an  example for the halo formation  rate when $\mmin$ is
    only one order of magnitude below the halo mass. }
  \label{figs:rate}
\end{figure}

The  upper  panel  of  Fig.~\ref{figs:rate} shows  the  star-formation
histories of elliptical galaxies as a function of the mass of the halo
in which these galaxies are located at $z=0$, from the semi-analytical
simulations  of \citet[][their  Figure~3]{DeLucia06}.  Note  that this
model predicts  that ellipticals in  more massive haloes  formed their
stars earlier, and over a shorter period of time, in good, qualitative
agreement  with   the  observational  data   of  \citet{Thomas05}  and
\citet{Nelan05}.   The lower panel  of Fig.~\ref{figs:rate}  shows the
corresponding formation rates of the dark matter haloes, as defined by
equation~(\ref{eq:mall_rate}).    Here  we   have  adopted   the  same
cosmological    parameters    as   in    De    Lucia   \etal    (i.e.,
$\Omega_{\Lambda}=0.75$,  $\Omega_m=0.25$,  $\sigma_8=0.9$,  $h=0.73$)
and   we  have   used  a   constant   threshold  mass   of  $\mmin   =
1.72\times10^{10} \hmsun$, corresponding to the mass resolution of the
numerical simulation used by these authors.

The  formation rates of  DM haloes,  as seen  in Fig.~\ref{figs:rate},
reveal a qualitatively similar ADS behavior as for elliptical galaxies
in  the  SAM  of  \citet{DeLucia06}  and  in  the  observational  data
\citep[e.g., ][]{Thomas05}.  Although  the agreement is extremely good
for  the  massive  haloes,  the  SAM predicts  a  significantly  later
formation  in  lower  mass  haloes,  indicating  that  the  downsizing
strength is larger  in the SAM.  This highlights  the crudeness of our
simplified  model for star  formation, while  assumes that  stars form
instantaneously  as  soon  as  the  halo mass  exceeds  $\mmin$.   The
comparison  with the  SAM  suggests  that this  is  a fairly  accurate
assumption  in  massive haloes.   In  low  mass  haloes, however,  the
baryonic feedback  processes modelled  in the SAM  must have  caused a
significant  delay  in  the  formation  of  the  stars.   Indeed,  the
efficiency of  supernova feedback to cause  such a delay  in larger in
lower mass  haloes \citep{Dekel86}. In  principle we can  increase the
downsizing strength for the dark  matter haloes by increasing $\mmin$.
For example, the thin solid line  plots the formation rates for a halo
of $10^{12} \msun$  but with a higher $\mmin$  of $10^{11}\msun$. This
brings the formation rates in  better agreement with the specific star
formation rates of elliptical galaxies in haloes of $10^{12} \msun$ in
the  SAM. Thus,  one may  mimic the  delays in  star formation  due to
supernova  feedback  effects by  an  increase  in  the star  formation
threshold mass $\mmin$, eventhough we do not necessarily consider this
very physical.  We conclude that the  ADS in galaxies  has its natural
origin in  the ADS of  $\mall$, while the baryonic  physics associated
with cooling,  star-formation, and  feedback merely causes  a shifting
and stretching  of the relative  formation histories.  The  main trend
with halo mass,  however, simply relates to the  dark matter formation
histories.

We define the mean formation epoch of a dark matter halo as
\begin{equation}
\label{eq:mean_wr_def}
\omega_R \equiv {\int_0^{t_0} R(\omega) \omega \dd t \over
\int_0^{t_0} R(\omega) \dd t}
\end{equation}
If,  for  simplicity,  we  keep  the  star  formation  threshold  mass
constant, i.e., $\mmin(z)=\mmin$, then this reduces to
\begin{equation}
\label{eq:mean_wr}
\omega_R = \omega_0 + \sqrt{ \frac{2}{\pi}
(S_{\rm min}-S_0) } \;\;\;.
\end{equation}
where we  have used the fact that the denominator  of
equation~(\ref{eq:mean_wr_def}) is equal to unity. Note that this mean
formation epoch is very similar to $\bwall$  of
equation~(\ref{eq:zfall}), but  with $\beta  \simeq 0.67$ replaced by
$\sqrt{2/\pi} \simeq 0.8$.

\begin{figure}
\centerline{ \hbox{ \epsfig{file=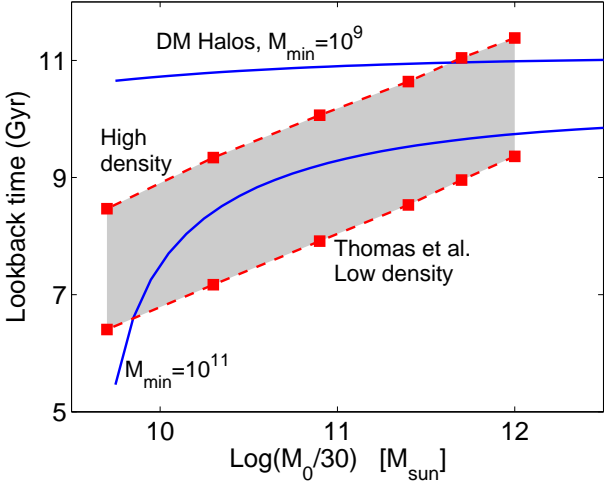,width=9cm} } }
  \caption{Effective formation epoch versus mass
    in EPS  theory versus observations.  The formation  epoch for dark
    matter    haloes    of    present    mass    $M_0$,    based    on
    equation~(\ref{eq:mean_wr}),  is  plotted  for $\mmin=10^{9}$  and
    $10^{11}\msun$ (solid  curves). Halo masses  are divided by  30 in
    order to roughly translate DM  into stellar masses.  The epoch for
    star   formation   as    deduced   from   local   ellipticals   by
    \citet{Thomas05}  is shown  (shaded area)  between the  two dashed
    lines  which refer  to  galaxies in  low-density and  high-density
    environments. Here  $h=0.75$ as  in \citet{Thomas05}.  There  is a
    downsizing behaviour in both cases.}
\label{figs:form_epoch}
\end{figure}

Figure~\ref{figs:form_epoch} compares  our analytic estimates  for the
mean  formation epoch  of DM  haloes to  the  star-formation histories
deduced from nearby elliptical  galaxies by \citet{Thomas05}, both for
ellipticals in  low-density and high-density environments\footnote{The
  density is  defined as  the number of  galaxies within a  one degree
  radius,  see   \citet{Thomas05}  for  details}.    The  solid  lines
correspond to our  estimates of equation~(\ref{eq:mean_wr}) for $\mmin
= 10^9$  and $10^{11}\msun$, as  indicated. Note that we  have divided
the dark  matter masses  by 30 to  obtain a  very rough proxy  for the
stellar  mass.  A comparison  with the  data of  Thomas \etal  is only
trully meaningful if (i) all gas  in haloes with $M > \mmin$ is turned
into stars instantaneously, and (ii) haloes host only one galaxy whose
stellar mass is equal to $M_0/30$. Although neither of these is likely
to be  correct, the data and  `model' are in  qualitative agreement in
that the more massive structures  have formed earlier, i.e., the model
shows ADS. If $M_{\rm min}=10^9  h^{-1} \msun$, the DS strength is too
weak, accross the  mass range of interest, compared  to the data. This
indicates that  the baryonic  physics needs to  delay and  or suppress
star formation relatively more in lower mass haloes. Alternatively, if
$M_{\rm min}$  is significantly larger ($\sim  10^{11} h^{-1} \msun$),
the DS strength at fixed halo mass is stronger, and there is less need
to delay  or suppress  star formation in  order to globally  match the
data. However, this requires a yet unknown physical mechanism that can
prevent star formation in all haloes below this mass limit.

\bigskip

Yet another way to view the  ADS aspect of halo formation histories is
via the  mean epoch at which a  halo of mass $M_0$  at time $\omega_0$
has a progenitor of mass  $M$. The number density, $\dd N(\omega)$, of
progenitors with  masses in  the interval $M$  to $M+\dd M$  at time
$\omega$  is  given  by equation~(\ref{eq:condprobM}).  Using it to
weight the averaging of $\omega$ we obtain
\begin{equation}
\label{eq:prog}
\omega_p \equiv \frac{\int \dd N(\omega) \omega
\dd\omega} {\int \dd N(\omega)\dd\omega} =
 \omega_0 + \sqrt{ \frac{2}{\pi}(S - S_0) } \;\;\;.
\end{equation}
This resembles $\omega_R$ in equation~(\ref{eq:mean_wr}), meaning that
the mean formation epoch for a given $\mmin$ is equivalent to the mean
epoch  for progenitors  to be  of a  given mass,  $M=\mmin$.   The ADS
behaviour is  apparent in equation~(\ref{eq:prog}) from  the fact that
$\omega_p$  increases  with  $M_0$   (via  $S_0$).   This  implies
that progenitors {\it of a given mass} appear earlier in the merger
tree of a  more massive  present-day  halo.   This is  similar  to the
result obtained  by \citet{Mouri06},  who  also argue  that downsizing
is  a natural prediction of hierarchical formation scenarios.

\section{Downsizing in Time}
\label{sec:TDS}

As   mentioned  in  \S\ref{sec:intro},   there  is   another  observed
downsizing effect,  different from the  archeological downsizing dealt
with  so  far,  which  refers   to  the  decrease  with  time  of  the
characteristic  mass of the  galaxies with  the highest  specific star
formation rates. We  show   here  that,  unlike  the   archeological
downsizing,  this downsizing in time (DST) is  not in general rooted
in the hierarchical clustering of dark matter haloes.  The dark haloes
show such an effect only if $\mmin$ is decreasing with time in a
sufficiently steep pace.

\begin{figure}
 \centerline{ \hbox{ \epsfig{file=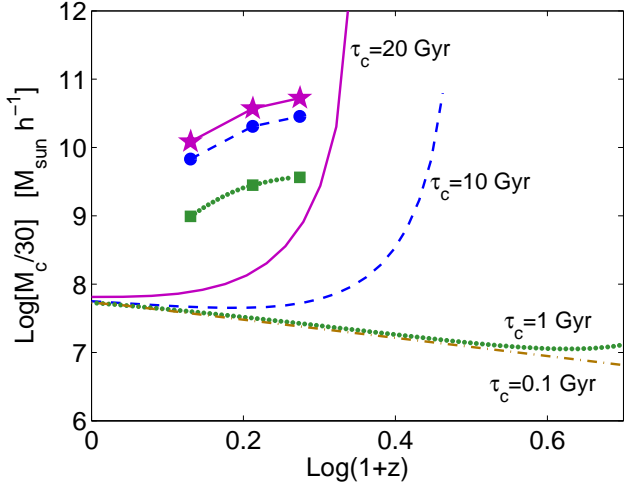,width=9cm} } }
  \caption{The mass of haloes which form at a given rate $R_c$ at $z$,
    $M_c(R=R_c,z)$  from   equation~(\ref{eq:DST}).   The  curves  are
    marked by  $\tau_c=R_c^{-1}$.  For high enough  $\tau_c$, the mass
    $M_c$ depends solely on $\dd\omega /\dd t$, while for low $\tau_c$
    it is given by $M_c(z)=\mmin(z)$.  The connecting symbols refer to
    the  observed star-formation time  scale for  galaxies of  a given
    stellar  mass   from  Fig.~3  of   \citet{Brinchmann00},  for  the
    corresponding values  of $\tau_c$.  The $M_c$ values  for the dark
    haloes were  divided by  30 in order  to allow a  crude comparison
    with the stellar mass of the galaxies.  }
  \label{figs:tds}
\end{figure}

The symbols in Fig~\ref{figs:tds} show  the DST obtained from the data
in \citet{Brinchmann00}.  For  a given SSFR we select  from their data
the stellar  mass and redshift of  a galaxy with that  SSFR. The solid
squares, connected by a dotted curve, plot the stellar mass of objects
forming their  stars with a SSFR  of 1 Gyr$^{-1}$,  corresponding to a
doubling time-scale  $\tau_c = 1$ Gyr. Note  that the characteristic
stellar mass of  systems forming stars at this rate  is lower at lower
redshift; this is DST. The  other symbols correspond to lower SSFRs of
0.1  Gyr$^{-1}$  (solid  dots  connected  by dashed  curve)  and  0.05
Gyr$^{-1}$ (stars connected by solid  curve).  Note that each of these
curves reveals DST, and that more massive systems have lower SSFRs, at
each redshift.

In  order to  compare  this with  dark  matter haloes,  we define  the
``current", specific formation rate of  dark matter haloes as the rate
of change of  $\mall$ normalized to $\mall$. This  rate is obtained by
setting    $\omega=\omega_0$   in    the   general    expression   for
$R(\omega\,\vert\, \omega_0,M_0)$ of equation~(\ref{eq:mall_rate}):
\begin{eqnarray}
 R(\omega \, \vert \, \omega,M_0) =
- \sqrt{\frac{2}{\pi}} \frac{1}{\sqrt{\smin(z)-S(M_0)}}
\frac{\dd\omega}{\dd t}(z) \,.
\label{eq:current_rate}
\end{eqnarray}
For  a  fixed  rate  $R=R_c$,  one can solve for $M_c(z)$, the mass of
haloes that are formed with the rate $R_c$:
\begin{equation}
\label{eq:m_rate} S\left[M_c(z)\right] = \smin(z) - \frac{2}{\pi}
\left( \frac{\dd \omega}{\dd t} \right)^2 \tau_c^{2} \,,
\label{eq:DST}
\end{equation}
where  $\tau_c\equiv R_c^{-1}$ is  the corresponding  time-scale.  The
curves  without  symbols   in  Fig~\ref{figs:tds}  show  the  $M_c(z)$
relations thus  obtained for  four different time-scales  $\tau_c$, as
indicated.  In order to allow for  a comparison with the data, we have
divided the halo masses by 30, as a rough proxy for stellar mass.  The
first thing  to notice is  that these `model predictions'  have almost
nothing in common  with the data.  First of all,  all $M_c(z)$ seem to
converge to  the same mass at  low $z$, independent  of $\tau_c$. This
owes  to the  fact  that  $R \rightarrow  \infty$  if $M_0
\rightarrow \mmin$;  the specific  formation  rate becomes  infinite
at $\mmin$. Secondly,  for  high  specific  formation rates  (low
$\tau_c$),  the $M_c(z)$ decreases with increasing $z$,  opposite to
the DST observed. This simply  owes to the fact  that low $\tau_c$
implies that $M_c(z) \sim \mmin(z)$, which, according to
equation~(\ref{eq:mmin}) decreases with  increasing   redshift.   When
$\tau_c$   is  sufficiently  high ($\gtrsim10$ Gyr), however, the dark
haloes show a qualitative DST, in that $M_c$ increases  with redshift.
This basically owes  to the fact that   the   contribution  from   the
$\dd\omega/\dd   t$  term   in equation~(\ref{eq:DST}) becomes
dominant over the term governed by the $\mmin(z)$  behavior.  We
conclude  that, in  general, the  formation histories of dark  matter
haloes do not show a  DST effect as observed for  galaxies.   It is
clear  that DST  must  be  driven by  baryonic processes, which must
strongly  decouple the star formation rates from the halo formation
rates.  The challenge  for the models will be to do so while
maintaining a fairly tight coupling at high $z$, which, as we have
shown, is required in order to explain the ADS.

\bigskip

Another measure  of DST  for DM  haloes is the  time evolution  of the
average halo  mass at which dark  matter is being  added to virialized
progenitors, namely
\begin{equation}
M_R= \frac{ \int R(\omega \, \vert \, \omega,M) M \frac{\dd n}{\dd M}
\dd M } { \int R(\omega \, \vert \, \omega,M) \frac{\dd n}{\dd M} \dd
M  } \,.
\end{equation}
Here $R$  is given by equation~(\ref{eq:current_rate})  and $\dd n/\dd
M$ is  the number density  of haloes per comoving  volume \citep[e.g.,
from][]{Sheth02}.

\begin{figure}
 \centerline{ \hbox{ \epsfig{file=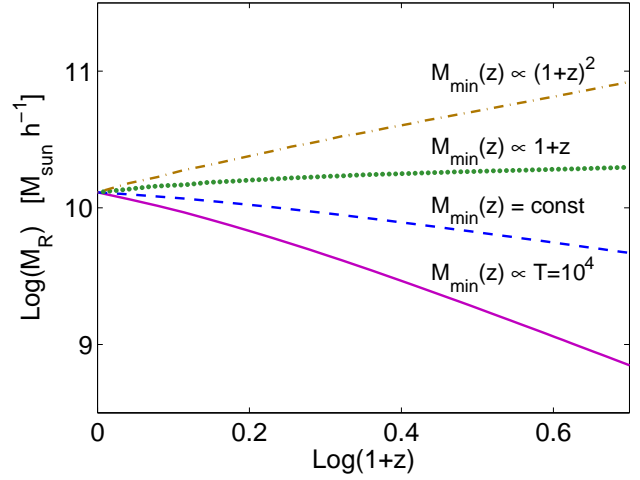,width=9cm} } }
  \caption{Average halo mass $M_R$ weighted by halo formation rate $R$
as a function of redshift for different $z$-dependences of $\mmin(z)$.
Downsizing in time is seen only when $\mmin(z)$ is increasing linearly
with $(1+z)$ or faster.}
  \label{figs:m_star_z}
\end{figure}

In Fig.~\ref{figs:m_star_z} we plot $M_R$ for several different growth
rates  of $\mmin(z)$,  all normalized  to coincide  with  our standard
value  of $\mmin$  at  $z=0$.   Similar to  $M_c$  defined above,  the
characteristic  mass $M_R$  is decreasing  with redshift  when  we set
$\mmin$ to correspond to a  constant virial temperature $T_{\rm vir} =
10^4 \K$  (see equation~[\ref{eq:mmin}]), or when  $\mmin$ is constant
in time.   Only when  $\mmin$ is increasing  with redshift  roughly as
$1+z$ or faster does the mass $M_R$ show a DST behaviour.


\section{Conclusions}
\label{sec:discuss}

We have  introduced a new  quantity to quantify  the growth of  a dark
matter halo merger tree, $\mall(z)$, the  sum of the masses of all the
virialized progenitors  at redshift  $z$ down to  a minimum  halo mass
$\mmin(z)$.   We have  shown,  using EPS  theory,  that this  quantity
reveals an ``archeological downsizing'' behavior in that $\mall(z)$ of
more  massive haloes grew  earlier and  on shorter  time-scales.  This
behaviour is present for any choice of non-zero $\mmin(z)$ and any
cosmology.  The only two conditions are (a) that the threshold mass
$\mmin(z)$  is independent of  the mass $M_0$ of  the present-day
halo,  and  (b) that  the  progenitor  mass function,  $P(M_1,z_1\vert
M_0,z_0)$    (equation.~[\ref{eq:probSS}])     either    depends    on
$S(M_0)-S(M_1)$   alone  (i.e.,   the  trajectories   $\delta(S)$  are
Markovian), or is such that  the fraction of mass in progenitors below
$M_{\rm min}$ decreases with increasing $M_0$. The fact that a similar
archeological downsizing effect is revealed by EPS  merger  trees and
in  N-body  simulations  indicates that  these conditions are at least
approximately valid.  One should note that the first condition,
although quite robust, might be  violated in certain circumstances.
For  example,  today's   halo  mass  $M_0$  could  be interpreted at
high $z$ as reflecting the  local environment density, and if the
threshold mass  is somehow affected by its environment this could
introduce a dependence of $\mmin(z)$ on $M_0$.

Using the  EPS formalism, we  have analytically formulated  the virial
mass  growth curve  $\mall(z)$, the  corresponding  formation redshift
$\zall$,  and  the  formation  rate.    The  latter  is  found  to  be
qualitatively  similar to the  formation rate  of stars  in elliptical
galaxies,  indicating that  the observed  archeological  downsizing in
these systems  has its  roots in the  formation histories of  the dark
matter haloes.  However, $\mall(z)$ is  only a good tracer of the star
formation histories  of galaxies if  all the gas  in haloes with  $M >
\mmin$ forms stars  instantaneously. In reality, this will  not be the
case,  as cooling  and  various feedback  processes  can delay  and/or
prevent the  formation of stars,  even in haloes  with $M \gg  \mmin$.
What  is  clear  from  our  study,  however, is  that  the  halo  mass
dependence of these baryonic processes has to be such that it does not
undo the mass dependence already encoded in $\mall(z)$.

We have also studied the  more common halo assembly histories, defined
as the  mass growth histories,  $\mbig(z)$, of the main  progenitor of
the merger tree. We have  developed an analytical approximation for it
based on EPS theory, and  confirmed the known ``upsizing" behaviour of
this assembly history.  We have  shown that it depends in principle on
the shape  of the  power-spectrum, but it  is valid for  all power-law
spectra as  well as  for the CDM  power spectrum. The  formation times
$\zbig$ and $\zall$, for a  sample of equal-mass haloes, were found to
be correlated  in a way  that can be  understood in terms of  the mass
growth at low redshifts.

The  downsizing  in  time, namely the decline with time of the mass of
star-forming  galaxies, cannot be easily traced back to the properties
of  the  dark matter  halo  merger trees. With our idealized recipe of
rapid star formation in virialized haloes above $\mmin(z)$, downsizing
in  time  can  be  reproduced only if $\mmin(z)$ is rapidly increasing
with  $z$.  Otherwise, this kind of downsizing is most likely a result
of   feedback  effects  on  star  formation,  which  requires  a  more
sophisticated  modeling  of the baryonic processes. The lesson is that
the different faces of ``downsizing'' reflect different phenomena, one
naturally  rooted  in  the hierarchical dark matter clustering process
and  the other determined by non-trivial baryonic processes, which are
yet to be properly modeled.

\section*{Acknowledgments}
We thank Risa Wechsler for providing the merger trees constructed from
cosmological  $N$-body simulations based on work supported by NASA ATP
NAG5-8218  (Primack and Dekel). We acknowledge stimulating discussions
with  Itai  Arad and Aaron Dutton. This research has been supported by
ISF  213/02  and  by  the  German-Israel  Einstein  Center  at  HU. AD
acknowledges  support  from  a  Blaise  Pascal  International Chair in
Paris.

\bibliographystyle{mn2e}
\bibliography{eyal}


\section*{Appendix: Analytical formulae for $\mbig$}

We use $\bmbig(z)$ to denote the main progenitor mass at redshift $z$,
averaged (at fixed $z$) over many individual merger trees for the same
parent  mass  $M_0$.  Here  we  use the  EPS  formalism  to derive  an
analytical estimate for $\bmbig(z)$.

\subsection*{Basic Equation}

Let's start with  a halo of mass $M_0$ at time  $\omega_0$, and take a
small time-step,  $\dW$, back in  time. At the time  $\omega_0+\dW$ we
want to compute the average mass of the main progenitor. This requires
the full  probability distribution, $P(M_{\rm main}  \vert M_0, \dW)$,
that a halo of mass $M_0$  at $\omega_0$ has a main progenitor of mass
$M_{\rm  main}$ at  time $\omega_0  +  \dW$.  For  $M_{\rm main}  \geq
M_0/2$, one has  that $P(M_{\rm main} \vert M_0,\dW)$  is equal to the
total    progenitor   distribution    $\dd   N/\dd    M$    given   by
equation~(\ref{eq:condprobM}),  simply  because  any progenitor  whose
mass exceeds  $M_0/2$ must be the \emph{main}  progenitor. For $M_{\rm
  main} <  M_0/2$, however, the only  valid condition is  that $P \leq
\dd N/\dd M$, which is not sufficient to predict $P(M_{\rm main} \vert
M_0,\dW)$.

As a first,  naive approximation we assume that  $P(M_{\rm main} \vert
M_0, \dW) = 0$ for $M_{\rm main} < M_0/2$, so that the main progenitor
always has  a mass $M_{\rm main}  \geq M_0/2$. Using  this
approximation,  the  average  mass  of  the  main progenitor,
$\bmbig(\dW)$, can be written as:
\begin{equation}
\label{eq:mmainaver} \bmbig(\dW) = \int_{M_0/2} ^{M_0} P(M\vert M_0,
\dW) M \dd M \,.
\end{equation}
Using the definition of $\dd N/\dd M$ this reduces to
\begin{equation}
\label{eq:mbig_dW}
\bmbig(\dW) =  M_0 \left[ 1 - {\rm
    erf}\left(\frac{\dW}{\sqrt{2S_2-2S_0}} \right) \right] \,,
\end{equation}
with $S_2=S(M_0/2)$ and $S_0=S(M_0)$.

We  assume that  $M_0$ is  just the  main progenitor  of  the previous
time-step\footnote{Equation~(\ref{eq:mbig_dW}) becomes linear in $\dW$
  for  small enough  $\dW$, and  this  gives: $\bmbig(\dW_1+\dW_2\vert
  M_0)  = \bmbig(\dW_2\vert  \bmbig(\dW_1\vert M_0$))}.   The  rate of
change, $\dd \bmbig /\dd\omega$, can then be computed as
\begin{eqnarray}
\frac{\dd \bmbig }{\dd\omega} & = & \lim_{\dW \rightarrow 0}
\frac{ \bmbig(\dW) - M_0 }{\dW} \\ \nonumber
& = & -M_0 \lim_{\dW \rightarrow 0} \frac{1}{\dW} \rm{erf}\left(
\frac{\dW}{\sqrt{2S_2-2S_0}} \right) \,.
\end{eqnarray}
Using that $\rm{erf}(x)  \rightarrow 2x/\sqrt{\pi}$ when $x\rightarrow
0$ this yields:
\begin{equation}
\label{m_asm} \frac{\dd\bmbig}{\dd\omega}
=-\sqrt{\frac{2}{\pi}}\frac{\bmbig}{\sqrt{S_2-S}} \,.
\end{equation}

In the case of scale  free initial conditions, were the power-spectrum
is a pure power law ($S\propto M^{-\alpha}$) we can solve for $\bmbig$
analytically:
\begin{equation}
\bmbig(\omega) = \left[ M_0 ^{-\frac{\alpha}{2}} + c_{\alpha}
(\omega-\omega_0) \right]^{-\frac{2}{\alpha}} \,,
\end{equation}
where  $c_{\alpha}=\alpha  \left[2\pi S(M=1) \left(2^{\alpha}-1\right)
\right]^{-1/2}$.

The above  derivation is based on  the assumption that the main
progenitor always has a mass $M_{\rm main} \geq M_0/2$ in the limit
$\dW \rightarrow 0$. However,  as we show below, when $\dW \rightarrow
0$ the probability  that $M_{\rm main}<M_0/2$ decrease like $\dW$.
Consequently, this will give a non-negligible effect for sufficiently
large $\omega$.

\subsection*{Towards better accuracy}

\begin{figure}
  \centerline{ \hbox{ \epsfig{file=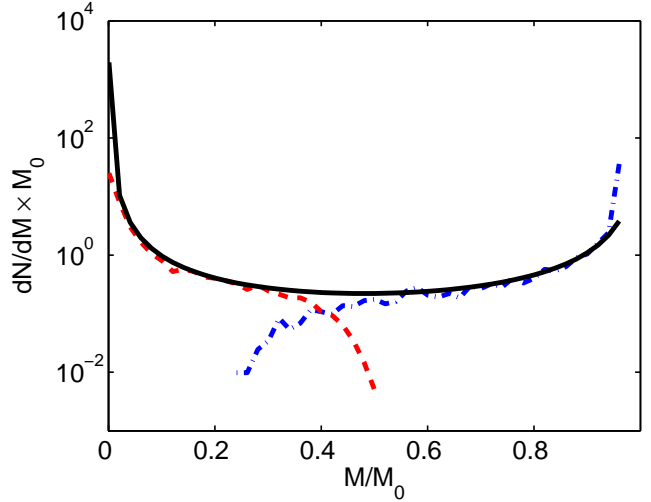,width=9cm} } }
  \caption{Average number of progenitors of mass $M$ one time-step $\dW=0.1$
    before the present for halo of mass $M_0=10^{12}\hmsun$. The solid
    curve  is  the   theoretical  prediction.   The  dot-dashed  lines
    indicates that  mass distribution of the  main progenitor (defined
    as  the  most  massive  progenitor),  as  obtained  from  $10.000$
    Monte-Carlo  EPS realizations  based on  the SK99  algorithm. Note
    that  the  probability  distribution  for  the mass  of  the  main
    progenitor, equals  $\dd N/\dd  M$ down to  $M_0/2$, but  does not
    vanish down to $M\sim 0.25M_0$. Finally, the dashed line indicates
    the mass distribution of the least massive progenitors obtained in
    these $10.000$ realizations.}
  \label{figs:big_distrib}
\end{figure}

The   dot-dashed  line   in   Fig.~\ref{figs:big_distrib}  shows   the
distribution of  the main progenitor masses  of a halo of  mass $M_0 =
10^{12}  \hmsun$  in  a  single  time step  $\dW=0.1$,  obtained  from
$10.000$ realizations  based on the  SK99 algorithm.  One  can clearly
see that  $P(M_{\rm main} \vert  M_0, \dW)$ has a  non-negligible tail
for $M_{\rm main}  < M_0/2$.  The following analysis  aims to find the
solution for $\bmbig$ taking this low-mass tail into account.

The  correct  shape  of  $P(M_{\rm  main}  \vert  M_0,  \dW)$  can  be
constrained by the following conditions:
\begin{enumerate}

\item The integral of $P(M_{\rm main} \vert M_0, \dW)$ over all masses
  should equal unity, for all time-steps $\dW$.

\item  $P(M_{\rm main} \vert  M_0, \dW)  = \dd  N/\dd M$,  for $M_{\rm
    main}  \geq M_0/2$  (equation~[\ref{eq:condprobM}]),  and  $P(M_{\rm
    main}  \vert M_0,  \dW) \leq  \dd N/\dd  M$, for  $M_{\rm  main} <
  M_0/2$

\item $P(M_{\rm main} \vert M_0, \dW)$ should not depend on the
  time-step subdivisions.  This can be written as:
\begin{eqnarray}
\lefteqn{P(M_{\rm main} \vert M_0, \dW_1 + \dW_2) =} \\ \nonumber
  & & \int P( M_1 \vert M_0, \dW_1)
P(M_{\rm main} \vert M_1, \dW_2) \dd M_1 \,.
\end{eqnarray}
\end{enumerate}
In what follows  we estimate the limits on  $P(M_{\rm main} \vert M_0,
\dW)$ using conditions (i) and (ii),  and show that they give a narrow
range for $\bmbig(z)$.  Condition (iii) does not force the solution to
be unique, hence it enables a  set of solutions, each of them is valid
within the  EPS formalism.  Because condition (iii)  is more difficult
to implement,  we do  not compute its  effect on $\bmbig$,  and assume
that it will not significantly affect the range of solutions.

The first  condition on $P(M_{\rm main}  \vert M_0, \dW)$  is that its
integral equals unity.  We define  $n_{\rm tail}$ as the integral over
$P(M_{\rm  main} \vert  M_0, \dW)$  from $M_{\rm  main}=0$  to $M_{\rm
  main} = M_0/2$:
\begin{equation}
\label{eq:ntail}
n_{\rm tail} = 1 - \frac{M_0}{\sqrt{2\pi}}\int_{S_0}
^{S_2} \frac{1}{M} \frac{\dW}{\Delta S^{1.5}} \exp\biggl[-
\frac{\dW^2}{2\Delta S} \biggr] \dd S \,,
\end{equation}
where $\Delta S=S-S_0 = S(M) - S(M_0)$.

We  can estimate  the possible  effect  \emph{any} tail  will have  on
$\bmbig$, by computing the effect  of the most extreme tails possible.
The first extreme is to concentrate all the tail in a small range near
$M=0$.  In  this case,  the integral of  $M_{\rm main}  P(M_{\rm main}
\vert M_0, \dW)$ over the range  $0 \leq M_{\rm main} < M_0/2$ will be
zero.  As a  result, the $\bmbig(z)$ that corresponds  to this extreme
is given  by equation~(\ref{eq:mbig_dW}).  The second extreme  is that
all the  tail is concentrated  near $M_0/2$, that is,  $P(M_{\rm main}
\vert M_0,  \dW)$ has its  maximum values ($=\dd  N/\dd M$) down  to a
lower mass  limit, $M/q_{\rm  max}$, which is  set by  the requirement
that
\begin{equation}
\label{eq:ntailt}
n_{\rm tail} = {M_0 \over \sqrt{2\pi}} \int_{S_2}
^{S_{q+}} \frac{1}{M} \frac{\dW}{\Delta S^{1.5}} \exp\biggl[-
\frac{\dW^2}{2\Delta S} \biggr] \dd S \,,
\end{equation}
where $S_{q+} = S(M/q_{\rm max})$. If we focus our attention on small
time-steps $\dW$, then we can use that $\dd P(M_{\rm main} \vert M_0,
\dW)/\dd M_{\rm main} \simeq 0$ near $M_0/2$ to approximately write
that
\begin{equation}
\label{eq:sqnul} S_{q+} \simeq S_2 + \sqrt{{\pi \over 2}} {(S_2 -
S_0)^{1.5} \over \dW} n_{\rm tail} \,.
\end{equation}
This enables us to use a simple equation for $S_{q+}$, combined with
the definition of $n_{\rm tail}$ in equation~(\ref{eq:ntail}).

What remains is  to find an appropriate expression  for $n_{\rm tail}$
which is  valid in the limit  of small time-steps  $\dW$.  We
therefore split the  integral in  equation~(\ref{eq:ntail}) into two
parts. The first one ($n_1$) is for  the range $0<\Delta S< \Delta
S_{\epsilon}$, where  $\dW  \ll  \Delta  S_{\epsilon}\ll  1$  and  we
can  make  the approximation $M\sim M_0$:
\begin{eqnarray}
\label{eq:none}
  n_1 &\simeq&  \frac{1}{\sqrt{2\pi}}\int_{S_0} ^{S_0+\Delta
S_{\epsilon}} \frac{\dW}{\Delta S^{1.5}} \exp\left[
\frac{-\dW^2}{2\Delta S} \right] \dd S \\
  \nonumber &=& 1 - \rm{erf}\left[ \frac{\dW}{\sqrt{2\Delta
S_{\epsilon}}} \right] \simeq 1 -
\sqrt{\frac{2}{\pi}}\frac{\dW}{\sqrt{\Delta S_{\epsilon}}} \,.
\end{eqnarray}
The  second  range ($n_2$) is for $\Delta S>\Delta S_{\epsilon}$ where
the  approximation  $\exp\left[-\dW^2/(2\Delta S) \right] \simeq 1$ is
valid:
\begin{equation}
\label{eq:ntwo}
n_2 \simeq \frac{M_0}{\sqrt{2\pi}}\int_{S_0+\Delta S_{\epsilon}}
^{S_2} \frac{1}{M} \frac{\dW}{\Delta S^{1.5}} \dd S \,.
\end{equation}
Combining equations~(\ref{eq:ntail}), (\ref{eq:none})
and~(\ref{eq:ntwo}) then yields :
\begin{equation}
n_{\rm tail} =  \sqrt{\frac{2}{\pi}}\dW\bigl[ \Delta
S_{\epsilon}^{-0.5} - \frac{M_0}{2}\int_{S_0+\Delta S_{\epsilon}}
^{S_2} \frac{1}{M} \frac{\dd S}{\Delta S^{1.5}} \bigr] \,.
\end{equation}
Finally we take the limit $\dS_{\epsilon} \rightarrow 0$, and obtain
\begin{equation}
n_{\rm tail} = \sqrt{\frac{2}{\pi}}\dW \left[
\frac{1}{2}\int_0^{S_2-S_0} \frac{M-M_0}{M}\frac{\dd \dS}{\dS^{1.5}}
+ \frac{1}{\sqrt{S_2-S_0}} \right] \,,
\end{equation}
Substitution in (\ref{eq:sqnul}) then yields
\begin{equation}
S_{q+} \simeq 2S_2 - S_0 + \frac{(S_2-S_0)^{1.5}}{2}\int_0^{S_2-S_0}
       \frac{M-M_0}{M}\frac{\dd \dS}{\dS^{1.5}} \,,
\end{equation}
independent of  $\dW$.  For  the standard $\Lambda$CDM  cosmology this
yields $2.1  < q_{\rm max}  < 2.3$ for  $0.1 < \Omega_m \leq  0.9$ and
$10^8 h^{-1} \msun \leq M_0  \leq 10^{15} h^{-1} \msun$. This implies
that although there is a negligible probability that the main
progenitor has a mass $M_0/2.3 < M_{\rm main} < M_0/2$ when $\dW
\rightarrow 0$, this probability behaves like $\dW$ and it cannot be
neglected. We can take this into account by rewriting
equation~(\ref{m_asm}) as
\begin{equation}
\label{m_asm_better}
\frac{\dd\bmbig}{\dd\omega} =
-\sqrt{\frac{2}{\pi}}\frac{\bmbig}{\sqrt{S_q-S}} \,.
\end{equation}
with $S_q = S(M_0/q)$ and $2 \leq q \lta 2.3$. In principle, any value
of $q$  in the  range above is  allowed.  In particular,  merger trees
constructed using  different algorithms  may have different  values of
$q$ in the above range, as long as the algorithms adopt a sufficiently
small time-step $\dW$.

\begin{figure}
 \centerline{ \hbox{ \epsfig{file=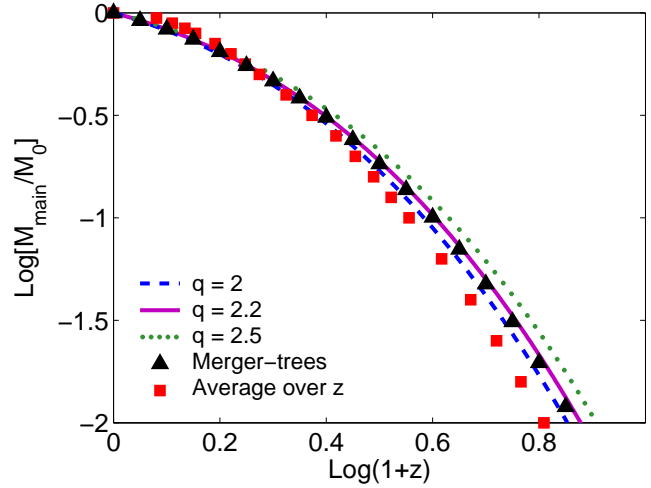,width=9cm} } }
  \caption{The mass of the main progenitor as computed in different ways
    for  $M_0=5\times10^{12} \hmsun$.   Big  circles are  from vdB02.
    Squares are the average of  EPS merger trees, averaged over $z$ at
    a  fixed  $M$.  Triangles  are  from  the  same merger  trees  but
    averaged  over  $M$  at  a  fixed  $z$.   Dashed  lines  show  the
    theoretical limits  of the analytic  formula ($q=2$ and  $q=2.5$).
    The solid line is the analytic formula with $q=2.2$.}
  \label{figs:mbig_app}
\end{figure}

In  Fig.~\ref{figs:mbig_app}  we show  $\bmbig$  for  a  halo of  mass
$5\times10^{12}\hmsun$.  The triangles  show the results obtained from
many  independent  EPS  merger   trees,  constructed  using  the  SK99
algorithm. Note that  the averaging is done over  $M_{\rm main}(z)$ at
fixed  $z$, which  is the  same as  done in  the  analytical estimates
(equation~[\ref{eq:mmainaver}]).  The  dashed, solid and  dotted lines
correspond       to      the       analytical       prediction      of
equation~(\ref{m_asm_better})   for   $q=2$,   $q=2.2$  and   $q=2.5$,
respectively. The curve for $q=2.2$ is in excellent agreement with the
EPS merger trees. Note that this  value for $q$ is within the expected
range.  In  Fig.~\ref{figs:mbig} we  show that our  analytical formula
with $q=2.2$  also accurately fits  the merger-tree results  for other
values of $M_0$.

\subsection*{A Universal Fitting Function}

Equation~(\ref{m_asm_better})  can  be  solved  to  obtain  a  direct,
analytical formula  for $\bmbig(z)$.  We use the  fitting function for
$S(M)$ given in vdB02:
\begin{equation}
S(M) = g^2 \biggl[ \frac{c\Gamma}{\Omega_m^{1/3}}M^{1/3} \biggr]
\cdot \frac{ \sigma_8^2 } {g^2(32\Gamma)} \,,
\end{equation}
where $c=3.804\times 10^{-4}$, and $g(x)$ is an analytical function:
\begin{eqnarray}
\lefteqn{ g(x) = 64.087 \bigl[ 1 + 1.074x^{0.3} } \\
\nonumber & &  - 1.581x^{0.4} + 0.954x^{0.5} - 0.185x^{0.6} \bigr] ^
{-10} \,.
\end{eqnarray}
In terms of the new set of variables
\begin{eqnarray}
  \widehat{M} = M \frac{\Gamma^3}{\Omega_m} \;\; , \;\;
  \widehat{\omega} = \omega \frac{g(32\Gamma)}{\sigma_8} \;\; , \;\;
  \widehat{g}(x) = g(c \cdot x^{1/3} ) \,.
\end{eqnarray}
equation~(\ref{m_asm_better}) does not depend on cosmology, and can be
easily solved to give
\begin{equation}
\widehat{M} = F_q^{-1}[ \widehat{\omega}-\widehat{\omega}_0 +
F_q(\widehat{M}_0) ] \,,
\end{equation}
where
\begin{equation}
F_q(\widehat{M})=-\sqrt{\frac{\pi}{2}} \int_0^{\widehat{M}}
\frac{\sqrt{\widehat{g}^2(\widehat{M'}/q)-\widehat{g}^2(\widehat{M'})}}
{\widehat{M'}} \dd\widehat{M'} \,.
\end{equation}
Finally we write the solution in the original variables:
\begin{equation}
\label{ma equation1}
\bmbig(\omega) = \frac{\Omega_m}{\Gamma^3}F_q^{-1} \left[
\frac{g(32\Gamma)}{\sigma_8}(\omega-\omega_0) + F_q\left( \frac{
\Gamma^3}{\Omega_m} M_0 \right) \right] \,.
\end{equation}
The analytical fitting function for $F_q(u)$ with $q=2.2$ is
\begin{eqnarray}
\lefteqn{F_q(u) = -6.92\times 10^{-5} \ln ^4 u +
5.0\times 10^{-3} \ln ^3 u +} \\ \nonumber
& & + 8.64\times10^{-2} \ln^2u - 12.66 \ln u + 110.8 \,,
\end{eqnarray}
which  is  accurate  to  better   than  one  percent  over  the  range
$1.6\times10^{4} < M\Gamma^3\Omega_m^{-1} < 1.6\times10^{13}\hmsun$

\subsection*{An Alternative Method}

We  now present an  alternative method  to compute  $M_{\rm main}(z)$,
which is based  on the method originally introduced  by LC93. The LC93
argument is  as follows: at a  specific redshift $z$,  one can compute
the  probability for  having  a  progenitor with  a  mass larger  than
$M_0/2$.   This probability equals  the probability  that a  tree will
have  its $\zbig$  greater  than  $z$ (where  $\zbig$  is defined  by:
$\mbig(\zbig) = M_0/2$).  Although LC93 claim their formula is only an
approximation,  we have not  found any  gap in  their argument,  so we
think this should be an accurate prediction.

Lets  define  $Q(\omega_1 \vert M_0,\omega_0)$ as the probability that
a  halo  with  mass $M_0$ at time $\omega_0$ will have its merger-tree
obey   $\omega(\zbig)=\omega_1$.   The   probability   for   having  a
progenitor with mass bigger than $M_0/2$ then equals :
\begin{equation}
\label{lc_int} \int_{S_0}^{S_2}
\frac{M_0}{M}f(S-S_0,\omega-\omega_0)\dd S = \int_{\omega}^{\infty}
Q(\omega_1 \vert M_0,\omega_0) \dd \omega_1 \,,
\end{equation}
where $f$  is defined in  equation~(\ref{eq:probS}), $S_0=S(M_0)$, and
$S_2=S(M_0/2)$.   The distribution  $Q(\omega \vert  M_0,\omega_0)$ is
obtained  by  differentiating  the  above  equation  with  respect  to
$\omega$  and multiplying  it by  -1.  In  order to  compute  the mean
formation  time, we  need  to average  $\omega$  over the  probability
distribution $Q$:
\begin{eqnarray}
\label{eq:bwbiga_int}
\lefteqn{ \bwbiga = \int_{\omega_0}^{\infty}
\omega Q(\omega \vert M_0,\omega_0) \dd \omega =} \\ & & \nonumber -
\int_{\omega_0}^{\infty} \omega\dd \omega \frac{\partial}{\partial
\omega} \int_0^{S_2-S_0} \frac{M_0}{M(S_0+\dS)}f(\dS,\dW)\dd\dS \,.
\end{eqnarray}
Here  $\bwbiga$ is  obtained  by averaging  over  all $\omega$  (time)
possible for getting a mass $M_0/2$. This is different from the method
used  above,  where  $\wbig$   was  computed  by  averaging  the  main
progenitor  masses at  a  fixed  time.  The  two  methods should  give
slightly  different  results, even  if  both  are  accurate.  This  is
illustrated in  Fig.~\ref{figs:mbig_app} where the  triangles indicate
the  average,  main  progenitor  history obtained  by  averaging  over
$\mbig$ at fixed $z$, while the squares show the results obtained when
averaging over $z$ at fixed $\mbig$.

So  far  we  have repeated  the  analysis  in  LC93. Now,  instead  of
computing  the  derivative  of  equation~(\ref{lc_int}),  we  simplify
equation~(\ref{eq:bwbiga_int}) by a simple integration by parts:
\begin{eqnarray}
\lefteqn{ \bwbiga = -\omega \left[ \int_0^{S_2-S_0}
\frac{M_0}{M}f(\dS,\dW)\dd\dS \right]_{\omega_0}^{\infty} +} \\
\nonumber & & + \int_{\omega_0}^{\infty} \dd \omega \int_0^{S_2-S_0}
\frac{M_0}{M(S_0+\dS)}f(\dS,\dW)\dd\dS \,.
\end{eqnarray}
The left  part is just  $\omega_0$\footnote{One should take  care when
  assigning the  integral lower limit. The integral  over $\dS$ should
  first   be   computed,   similar    to   the   analysis   done   for
  equation~(\ref{eq:ntail}).}.  We  can switch  the  integrals on  the
right  hand  side of  the  equation,  and  compute the  integral  over
$\omega$ first.  Finally we have:
\begin{equation}
\label{eq:wbiga}
\bwbiga-\omega_0 = \frac{M_0}{\sqrt{2\pi}}\int_0^{S_2-S_0}
\frac{\dd\dS}{M(S_0+\dS)\sqrt{\dS}} \,.
\end{equation}

For a self-similar power spectrum with $S\propto M^{-\alpha}$ the
integral can be done analytically:
\begin{equation}
\bwbiga-\omega_0 = \sqrt{\frac{2}{\pi}(S_2-S_0)} \;
_2F_1\left(\frac{1}{2},-\frac{1}{\alpha},\frac{3}{2},1-2^{\alpha}
\right) \,,
\end{equation}
where $_2F_1$  is the Gauss  Hypergeometric function. We can  see that
$(\bwbiga-\omega_0)/\sqrt{S_2-S_0}$ has the same value for all masses,
and thus it is  the natural variable to choose (as was  done in LC93).
On  the  other  hand,  we  showed  in  equation~(\ref{eq:wmain})  that
$(\bwbig-\omega_0)/(\sqrt{S_2}-\sqrt{S_0})$  is a also  constant, when
the averaging of the trees is made along the mass axis.

The  analysis  above  is  still   valid,  if  we  replace  $S_2$  with
$S(\bmbiga)$, so that  we can easily generalize this  result to obtain
$\bmbiga$  in the  range $\bmbiga>M_0/2$.   For masses  below $M_0/2$,
however,  we   cannot  compute  $\bmbiga$  since   this  requires  the
probability  for  getting  a  main  progenitor with  mass  lower  than
$M_0/2$.  As discussed above, this part of the probability function is
unknown.

\begin{figure}
 \centerline{ \hbox{ \epsfig{file=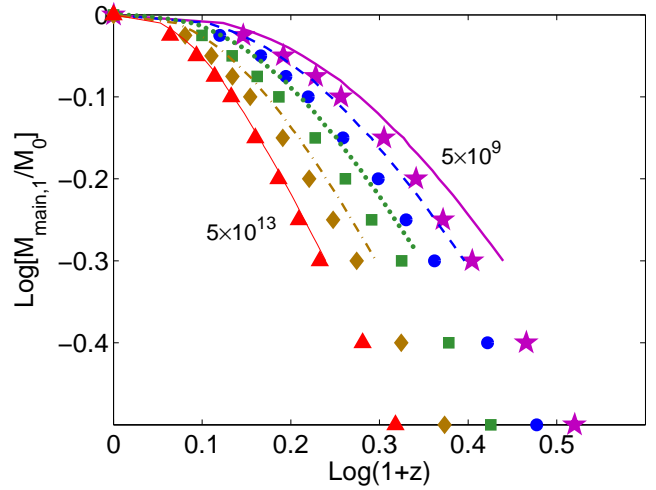,width=9cm} } }
  \caption{ $\bmbiga(z)$ for several halo masses. Symbols are
    from  the  merger trees  of  vdB02.  The  halo masses  range  from
    $5\times10^9$ to  $5\times10^{13} h^{-1}\;M_{\odot}$, spaced  by a
    decade.  Smoothed  lines are the results of  the analytic formula,
    Eq.~\ref{eq:wbiga},   replacing  $S_2$   with   $S(\bmbiga)$.  The
    equation is valid only for $\bmbiga > M_0/2$.}
  \label{figs:w_lc93}
\end{figure}

In  Fig.~\ref{figs:w_lc93} we  compare  results from  EPS merger  tree
realization  (vdB02, plotted  as  symbols) to  the analytical  formula
(smoothed lines). There are some deviation between the two, presumably
because the  averaging is done over  a large range  in redshift, where
the SK99  algorithm may have  slight inaccuracies. This effect  can be
seen  in  Fig.~\ref{figs:z_form_scatter} and  in  vdB02 (Fig.~4):  the
distribution of  formation times  is only accurate  for low  values of
$\wbig$.  Our previous method  for computing $\bwbig$ was not affected
by this  inaccuracy because  it was derived  using the limit  of small
time-steps behaviour.   This, combined with the fact  that this method
can only be used to compute  $\bmbig(z)$ down to $M_0/2$, and the fact
that  equation~(\ref{eq:wbiga})  cannot  be generalized  easily  since
$S_0$  is  buried inside  the  integrand,  clearly  favors the  method
discussed at  the beginning  of this appendix  over the  one discussed
here.

\label{lastpage}

\end{document}